\pdfoutput=1

\documentclass[11pt,twoside,a4paper,cmspaper,final,collab]{cms-tdr}
\def\svnVersion{ea080f0-D}\def\svnDate{2020/03/26}\def\cmsCernNoTag{CERN-EP-2019-213}\def\cmsCernDate{\today}\def\cmsMessage{"Published in Physical Review D as \href{http://dx.doi.org/10.1103/PhysRevD.101.052010}{\doi{10.1103/PhysRevD.101.052010}.}"}

\begin{document}\cmsNoteHeader{SUS-19-007}

\hyphenation{had-ron-i-za-tion}
\hyphenation{cal-or-i-me-ter}
\hyphenation{de-vices}
\RCS$Revision: 420218 $
\RCS$HeadURL: svn+ssh://svn.cern.ch/reps/tdr2/papers/SUS-19-007/trunk/SUS-19-007.tex $
\RCS$Id: SUS-19-007.tex 420218 2017-08-08 22:20:55Z richman $
\newlength\cmsFigWidth
\ifthenelse{\boolean{cms@external}}{\setlength\cmsFigWidth{0.98\columnwidth}}{\setlength\cmsFigWidth{0.6\textwidth}}
\ifthenelse{\boolean{cms@external}}{\providecommand{\cmsLeft}{upper\xspace}}{\providecommand{\cmsLeft}{left\xspace}}
\ifthenelse{\boolean{cms@external}}{\providecommand{\cmsRight}{lower\xspace}}{\providecommand{\cmsRight}{right\xspace}}
\ifthenelse{\boolean{cms@external}}{\providecommand{\cmsTable}[1]{#1}}{\providecommand{\cmsTable}[1]{\resizebox{0.92\textwidth}{!}{#1}}}

\ifthenelse{\boolean{cms@external}}{\providecommand{\CL}{\ensuremath{\text{C.L.}}\xspace}}{\providecommand{\CL}{\ensuremath{\mathrm{CL}}\xspace}}
\newlength\cmsTabSkip\setlength{\cmsTabSkip}{1ex}

\newcommand{\largr}{large-\ensuremath{R}\xspace}
\newcommand{\smalr}{small-\ensuremath{R}\xspace}

\newcommand{\ptvecell}{\ensuremath{\ptvec^{\kern2pt\ell}}\xspace}
\newcommand{\highmj}{high-\ensuremath{M_J}\xspace}
\newcommand{\lowmj}{low-\ensuremath{M_J}\xspace}
\newcommand{\MJ}{\ensuremath{M_J}\xspace}
\newcommand{\njets}{\ensuremath{N_{\text{jets}}}\xspace}
\newcommand{\nb}{\ensuremath{N_{\text{b}}}\xspace}

\newcommand{\ST}{\ensuremath{S_{\mathrm{T}}}\xspace}

\newcommand{\mGlu}{\ensuremath{m(\sGlu)}\xspace}
\newcommand{\mLSP}{\ensuremath{m(\PSGczDo)}\xspace}
\newcommand{\zjets}{{{\cPZ}\ensuremath{+}jets}\xspace}
\newcommand{\wjets}{{{\PW}\ensuremath{+}jets}\xspace}
\newcommand{\ttjets}{{{\ttbar}\ensuremath{+}jets}\xspace}

\hyphenation{had-ron-i-za-tion}
\hyphenation{cal-or-i-me-ter}
\hyphenation{de-vices}

\cmsNoteHeader{SUS-19-007}
\title{Search for supersymmetry in pp collisions at \texorpdfstring{$\sqrt{s}=13\TeV$ with 137\fbinv}{sqrt(s)=13 TeV with 137 inverse femtobarns} in
final states with a single lepton using the sum of masses of large-radius jets}

\date{\today}

\abstract{Results are reported from a search for new physics beyond the standard model in proton-proton collisions in final states with a single lepton; multiple jets, including at least one jet tagged as originating from the hadronization of a bottom quark; and large missing transverse momentum. The search uses a sample of proton-proton collision data at $\sqrt{s}=13\TeV$, corresponding to 137\fbinv, recorded by the CMS experiment at the LHC. The signal region is divided into categories characterized by the total number of jets, the number of bottom quark jets, the missing transverse momentum, and the sum of masses of large-radius jets.  The observed event yields in the signal regions are consistent with estimates of standard model backgrounds based on event yields in the control regions. The results are interpreted in the context of simplified models of supersymmetry involving gluino pair production in which each gluino decays into a top quark-antiquark pair and a stable, unobserved neutralino, which generates missing transverse momentum in the event. Scenarios with gluino masses up to about 2150\GeV are excluded at $95\%$ confidence level (or more) for neutralino masses up to 700\GeV. The highest excluded neutralino mass is about 1250\GeV, which holds for gluino masses around 1850\GeV.}

\hypersetup{%
pdfauthor={CMS Collaboration},%
pdftitle={Search for supersymmetry in pp collisions at sqrt(s)=13 TeV with 137 fb-1 in final states with a single lepton using the sum
of masses of large-radius jets},%
pdfsubject={CMS},%
pdfkeywords={CMS, physics, supersymmetry}}

\maketitle

\section{Introduction}
\label{sec:intro}
The physics program of the CMS experiment at the CERN LHC~\cite{1748-0221-3-08-S08001}
is designed to explore the TeV energy scale and to search for new particles and
phenomena beyond the standard model (SM), for example, those predicted by
supersymmetry (SUSY)~\cite{Ramond:1971gb,Golfand:1971iw,Neveu:1971rx,Volkov:1972jx,Wess:1973kz,Wess:1974tw,Fayet:1974pd,Nilles:1983ge}.
The search described here focuses on an important experimental signature that is
also strongly motivated by SUSY phenomenology. This signature includes
a single lepton (an electron or a muon);
several jets, arising from the hadronization of energetic quarks and gluons;
at least one \PQb-tagged jet, indicative of processes involving third-generation quarks;
and \ptvecmiss, the missing momentum in the direction transverse to the beam.
A large value of $\ptmiss \equiv \abs{\ptvecmiss}$
can arise from the production of high-momentum, weakly interacting
particles that escape detection. Searches for SUSY in the single-lepton final state have been performed by both
ATLAS and CMS in proton-proton ($\Pp\Pp$) collisions at $\sqrt{s}=7$, 8,
and 13\TeV~\cite{ATLAS:2011ad,Aad:2015mia,Chatrchyan:2011qs,Chatrchyan:2013iqa, Aad:2016eki,Aad:2016qqk,Aaboud:2017hrg,Khachatryan:2016uwr,Khachatryan:2016xdt,Sirunyan:2017fsj}.

This paper describes the single-lepton SUSY search based on the mass variable \MJ,
defined as the sum of the masses of large-radius jets in the event,
as well as on several other kinematic variables.
The search uses the combined CMS Run 2 data sample from 2016, 2017, and 2018,
corresponding to a total integrated luminosity of
approximately 137\fbinv at $\sqrt{s} = 13\TeV$.
The analysis is based on well-tested methods described in detail in two published
studies~\cite{Khachatryan:2016uwr,Sirunyan:2017fsj}, which used Run 2 data samples
of 2.3\fbinv (2015) and 35.9\fbinv (2016). In this version of the analysis,
the signal and control region definitions have been reoptimized to take advantage of the significant
increase in the size of the data sample and to maximize the analysis sensitivity to the SUSY models considered.
Other improvements, such as a more advanced \PQb tagging algorithm, have also been incorporated into the
analysis.

In models based on SUSY, new particles are introduced such that all fermionic
(bosonic) degrees of freedom in the SM are paired with corresponding bosonic (fermionic)
degrees of freedom in the extended, supersymmetric theory. The discovery of a Higgs boson at a low
mass~\cite{Aad:2012tfa,Chatrchyan:2012ufa,Chatrchyan:2013lba,Khachatryan:2014jba,Aad:2014aba,Aad:2015zhl}
highlighted a key motivation for SUSY, referred to as the gauge hierarchy
problem~\cite{tHooft:1979bh,Witten:1981nf,Dine:1981za,Dimopoulos:1981au,Dimopoulos:1981zb,Kaul:1981hi}.
Assuming that the Higgs boson is a fundamental (noncomposite) spin-0 particle,
its mass is subject to large quantum loop corrections, which would drive
the mass up to the cutoff scale of validity of the theory. If the SM is valid
up to the Planck scale associated with quantum gravity, these corrections
would be enormous. Stabilizing the Higgs boson mass at a low value, without invoking extreme
fine tuning of parameters to cancel the corrections, is a major theoretical challenge, which
can be addressed in so-called natural SUSY models~\cite{1988NuPhB.306...63B,Dimopoulos:1995mi,Barbieri:2009ev,Papucci:2011wy,Feng:2013pwa}.
In such models, several of the SUSY particles are constrained to be light~\cite{Papucci:2011wy}: the
top squarks, $\PSQt_{\mathrm{L}}$
and $\PSQt_{\mathrm{R}}$, which have the same electroweak gauge couplings as the left- (L) and right- (R) handed top
quarks, respectively; the bottom squark with L-handed couplings ($\PSQb_{\mathrm{L}}$); the gluino (\PSg); and the
Higgsinos (\sHig). In SUSY models that conserve $R$-parity---a multiplicative quantum number
equal to $+1$ for SM particles and $-1$ for their SUSY partners~\cite{Farrar:1978xj,Martin:1997ns}---SUSY particles
must be produced in pairs and each SUSY particle decay chain must terminate in the production of the lightest
supersymmetric particle (LSP). The LSP is therefore stable and,
if weakly interacting, can in principle account
for some or all of the astrophysical dark matter~\cite{Zwicky:1933gu,Rubin:1970zza,PDG2018}.

Motivated by the naturalness-based expectations that both the gluino
and the top squark should be relatively light, we search for
gluino pair production with decays to either off- or on-mass-shell top squarks.
Furthermore, gluino pair production has a large cross section relative
to most other SUSY pair-production processes, for a fixed SUSY particle mass.
Each gluino is assumed to decay
via the process $\PSg\to\PSQt_1\cPaqt$ (or the conjugate final state), where
the top squark mass eigenstate, $\PSQt_1$, is the lighter of the
two physical superpositions of $\PSQt_{\mathrm{L}}$ and $\PSQt_{\mathrm{R}}$.
Depending on the mass spectrum of the model, the top squark can
be produced either on or off mass shell, and it
is assumed to decay with 100\% branching fraction
via $\PSQt_1\to\cPqt\PSGczDo$, where \PSGczDo is a neutralino LSP. The neutralino is
an electrically neutral mixture of the neutral Higgsinos and electroweak gauginos.
Because the \PSGczDo is weakly interacting, it would traverse the detector without depositing energy, much
like a neutrino. As a consequence, neutralino production typically generates an apparent imbalance in
the total transverse momentum of the event, \ptvecmiss,
which is a priori known to be essentially zero, apart from detector resolution effects and missing momentum carried by weakly interacting
particles (\eg, neutrinos) or particles outside the detector acceptance.

\begin{figure*}[tp!]
\centering
\includegraphics[width=0.4\textwidth]{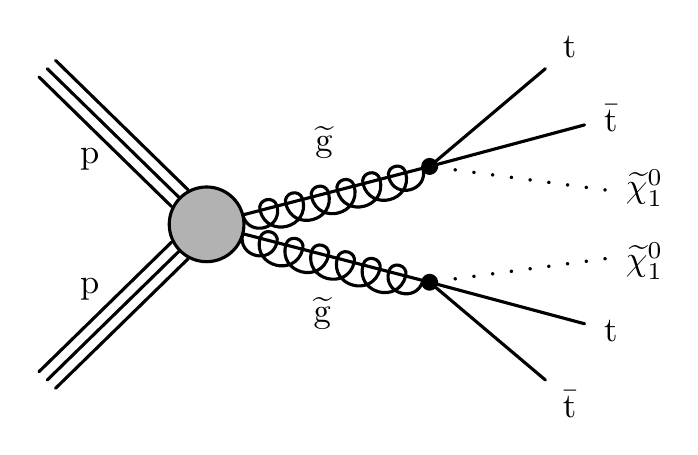}\hspace{0.1\textwidth}
\includegraphics[width=0.4\textwidth]{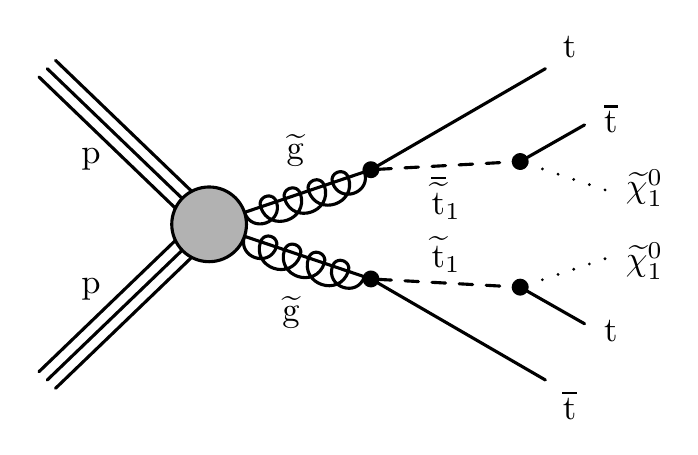}
\caption{Gluino pair production and decay for the simplified models T1tttt (left) and T5tttt (right). In T1tttt,
the gluino undergoes a three-body decay $\sGlu\to\ttbar\PSGczDo$ via a virtual intermediate top squark.
In T5tttt, the gluino decays via the sequential two-body process $\sGlu\to\PSQt_1\cPaqt$, $\PSQt_1\to\cPqt\PSGczDo$.
Because gluinos are Majorana fermions, each one can decay to $\PSQt_1\cPaqt$ and to
the charge conjugate final state $\smash{\PASQt_1\cPqt}$.
The filled circle represents the sum of processes that can lead to gluino pair production.
}
\label{fig:intro:nlep_t1tttt}
\end{figure*}

Diagrams showing gluino pair production with decays to off-mass-shell and on-mass-shell top squarks
are shown in Fig.~\ref{fig:intro:nlep_t1tttt} and are denoted as T1tttt and T5tttt, respectively,
in the context of simplified models~\cite{Chatrchyan:2013sza,bib-sms-2,bib-sms-3,bib-sms-4}.
Such models, which include only a small subset of the full SUSY particle spectrum,
are often used to quantify the results of searches,
in spite of limitations in describing the potential complexities associated with a complete spectrum.
The diagram for the T1tttt model does not explicitly show the off-mass-shell top squark, but the
fundamental gluino decay vertex for both T1tttt and T5tttt models is the same.
Thus, regardless of whether the top squark is produced on or
off mass shell, each gluino ultimately decays via the process $\PSg\to\ttbar\PSGczDo$, so signal events
would contain a total of four top quarks and two neutralinos.

The final states for both T1tttt and T5tttt are characterized by a large number of jets, four of which
are \PQb jets from top quark decays.  Depending on the decay modes of the accompanying \PW bosons, a range of
lepton multiplicities is possible. We focus here on the single-lepton final state, where the lepton is either
an electron or a muon, and a background estimation method specifically designed for this final state is
a critical part of the analysis. Events from the extreme tails of the kinematic distributions for \ttbar events
can have properties that closely resemble those of signal events, including
the presence of large \ptmiss generated by the neutrino
from a leptonic \PW boson decay. Initial-state radiation (ISR) from strong interaction
processes can enhance the jet multiplicity, producing another characteristic feature
of signal events. Quantifying the effects of ISR is a critical element
of the analysis.

The signature used here to search for the processes shown in Fig.~\ref{fig:intro:nlep_t1tttt} is characterized
not only by the presence of an isolated high transverse momentum (\pt) lepton,
multiple jets, at least one \PQb-tagged
jet, and large \ptmiss, but also by additional kinematic
variables. The first of these is \mT, defined as the transverse mass of the system consisting of the
lepton and the \ptvecmiss in the event. Apart from resolution and small
effects from off-mass-shell \PW boson production, \mT is bounded above by $m_{\PW}$
for events with a single leptonically decaying \PW boson, and this variable
is effective in suppressing the otherwise dominant single-lepton
\ttbar background, as well as background from \wjets events.

Largely because of the effectiveness of the \mT variable in helping to suppress
the single-lepton \ttbar background, the residual background in the signal regions
arises predominantly from a single SM process, dilepton \ttbar production. In such background events, both \PW bosons from the $\cPqt\to\cPqb\PW$ decays produce leptons, but only one of the two leptons satisfies the
lepton-identification criteria, as well as the requirements on the \pt, pseudorapidity ($\eta$),
and isolation from other energetic particles in the event. This background includes \ttbar events
in which one or both of the \PW bosons decay into a \Pgt lepton and its neutrino, provided that
the subsequent \Pgt decays produce a final state containing exactly one electron
or muon satisfying the lepton selection requirements.

A second key kinematic variable, \MJ, is defined as the
scalar sum of the masses of large-radius jets in the event. This quantity
is used both to characterize the mass scale of the event,
providing discrimination between signal and background, and as a key part of the
background estimation.  A property of \MJ exploited in this analysis is
that, for \ttbar events with large jet multiplicity,
this variable is nearly uncorrelated with \mT. As a consequence,
the \MJ background shape at high \mT, which includes the signal region, can be measured to
a very good approximation using the corresponding \MJ shape in a low-\mT control sample.
The quantity \MJ was first discussed in phenomenological studies, for example, in
Refs.~\cite{Hook:2012fd,Cohen:2012yc,Hedri:2013pvl}.
Similar variables have been used by ATLAS for SUSY searches in all-hadronic final states using 8\TeV
data~\cite{Aad:2015lea,Aad:2013wta}. Studies
of basic \MJ properties and performance in CMS have been presented
using early 13\TeV data~\cite{SUSYDPS}.

Because the signal processes would populate
regions of extreme tails of the SM distributions,
it is important to determine the background in a manner that
accounts for features of the detector behavior and
of the SM backgrounds that may not be perfectly modeled
in the simulated (Monte Carlo, MC) event samples. The background estimation
method is constructed such that corrections derived from MC
samples enter only at the level of double ratios of event yields,
rather than as single ratios. This approach helps to control the impact of potential
mismodeling on the background prediction because of the cancellation of many mismodeling effects.
Systematic uncertainties in these predicted double ratios are obtained by performing tests
using data control samples in regions that are kinematically similar but have only
a very small potential contribution from signal events.

This paper is organized as follows.
Sections~\ref{sec:eventSamples} and \ref{sec:CMSdetector} describe
the simulated event samples and the CMS detector, respectively.
Section~\ref{sec:triggerEventReconstruction} discusses
the triggers used to collect the data, the event reconstruction methods, and the
definitions of key quantities used in the analysis.
The event selection and analysis regions are presented in Section~\ref{sec:EvtSelAndRegions},
and the methodology used to predict the SM background is presented in Section~\ref{sec:backgroundEstimation}.
Section~\ref{sec:sysUncert} summarizes the systematic uncertainties
in the background predictions.
Section~\ref{sec:ResultsAndInterpretation} presents
the event yields observed in the signal regions,
the corresponding background predictions, the uncertainties associated with the
signal efficiencies, and the resulting exclusion regions
for the gluino pair-production models considered.
Finally, the main results are summarized in Section~\ref{sec:Summary}.

\section{Simulated event samples}
\label{sec:eventSamples}
The analysis makes use of several simulated event samples for modeling the signal and SM background processes.
These samples are used in the overall design and optimization of the analysis procedures, in the
determination of the efficiency for observing signal events, and in the calculation
of double-ratio correction factors, typically near unity, that are used in conjunction with event yields
in control regions in data to estimate the backgrounds in the signal regions.

The production of \ttjets, \wjets, \zjets, and quantum chromodynamics (QCD) multijet events is simulated with the MC
generator \MGvATNLO~2.2.2~\cite{Alwall:2014hca} in leading-order (LO) mode for 2016 samples and \MGvATNLO~2.4.2 for 2017 and 2018 samples. Single top quark
events are modeled with \MGvATNLO at next-to-leading order (NLO) for the \PQs-channel and with
\POWHEG~v2~\cite{Alioli:2009je,Re:2010bp} for the \PQt-channel and for associated $\cPqt\PW$ production.
Additional small backgrounds, such as \ttbar production in association with bosons, diboson processes, and
$\ttbar\ttbar$, are similarly produced at NLO with either \MGvATNLO or \POWHEG.
The events are generated using the NNPDF~2.3~\cite{Ball:2014uwa} set of parton distribution functions (PDF) for 2016 samples and the NNPDF~3.1~\cite{Ball:2017nwa} PDF set for 2017 and 2018 samples.
Parton showering and fragmentation are performed with the \PYTHIA~8.2~\cite{Sjostrand:2014zea} generator using
the CUETP8M1~\cite{Khachatryan:2110213} underlying event model for the 2016 samples and the CP5~\cite{Sirunyan:2019dfx} model for the 2017 and 2018 samples. The
detector simulation is performed with \GEANTfour~\cite{Agostinelli:2002hh}. The cross sections used to scale
simulated event yields are based on the highest order calculation
available.

Signal events for the T1tttt and T5tttt
simplified SUSY models are generated in a manner similar to that for the SM
backgrounds, with the
\MGvATNLO~2.4.2 generator in LO mode using the NNPDF~2.3 PDF set for 2016 samples and the NNPDF~3.1 PDF set for the 2017 and 2018 samples. Parton showering and fragmentation are performed with the
\PYTHIA~8.2 generator using the CUETP8M1~\cite{Khachatryan:2110213} underlying event model for the 2016 samples and the CP2~\cite{Sirunyan:2019dfx} model for the 2017 and 2018 samples.
However, because of the large number of model scenarios that must be considered,
the detector simulation is performed with the CMS fast simulation package~\cite{Abdullin:2011zz, Giammanco:2014bza},
with scale factors applied to account for differences with respect to the full simulation.
Event samples are generated for a representative set of model scenarios by scanning over the
relevant mass ranges for the \sGlu and \PSGczDo, and the yields are normalized to the cross-section at approximate
next-to-NLO, including next-to-next-to-leading-logarithmic (NNLL) 
contributions~\cite{Borschensky:2014cia,Beenakker:1996ch,Kulesza:2008jb,Kulesza:2009kq,Beenakker:2009ha,Beenakker:2011fu, Beenakker:2013mva,Beenakker:2014sma,Beenakker:2016lwe}. The modeling of the event kinematics is further
improved by reweighting the distribution of the number of ISR
jets to match the data based on a measurement in a dilepton \ttbar sample with two
\PQb-tagged jets~\cite{Sirunyan:2019ctn}.

Throughout this paper, two T1tttt benchmark models are used to illustrate typical signal behavior.  The
T1tttt(2100,100) model, with masses $\mGlu=2100\GeV$ and $\mLSP=100\GeV$, corresponds to a scenario with a
large mass splitting between the gluino and the neutralino. This mass
combination probes the sensitivity of the analysis to a low cross section (0.59\unit{fb}) process that has a
hard \ptmiss distribution, which results in a relatively high signal efficiency. The T1tttt(1900,1250) model, with
masses $\mGlu=1900\GeV$ and $\mLSP=1250\GeV$, corresponds to a scenario with a relatively small mass splitting (referred
to as a compressed spectrum) between the gluino and the neutralino. Here the cross section is higher (1.6\unit{fb})
because the gluino mass is lower than for the T1tttt(2100,100) model, but the sensitivity suffers from a low
signal efficiency due to the soft \ptmiss distribution.

Finally, to model the presence of additional $\Pp\Pp$ collisions from the same or adjacent bunch crossing
as the primary hard scattering process (pileup interactions), the simulated events are overlaid with
multiple minimum bias events (generated with the \PYTHIA~8.2 generator), such that the minimum
bias event multiplicity in simulation matches that observed in data.

\section{CMS detector}
\label{sec:CMSdetector}
The central feature of the CMS detector is a superconducting solenoid of 6\unit{m} internal diameter,
providing a magnetic field of 3.8\unit{T}. Within the solenoid volume are a silicon pixel and strip tracker,
a lead tungstate crystal electromagnetic calorimeter, and a brass and scintillator hadron calorimeter.
Each of these systems is composed of a barrel and two endcap sections. The tracking detectors cover the pseudorapidity range
$\abs{\eta}<2.5$. For the electromagnetic and hadronic calorimeters, the barrel and endcap detectors together
cover the range $\abs{\eta}<3.0$. Forward calorimeters extend the coverage to $3.0<\abs{\eta}<5.0$.
Muons are measured and identified in both barrel and endcap systems, which together cover
the pseudorapidity range $\abs{\eta} < 2.4$. The detection planes are based on three technologies:
drift tubes, cathode strip chambers, and resistive plate chambers, which are embedded in the steel
flux-return yoke outside the solenoid. The detector is nearly hermetic,
permitting accurate measurements of~\ptvecmiss.
Events of interest are selected using a two-tiered trigger system~\cite{Khachatryan:2016bia}. The first level (L1),
composed of custom hardware processors, uses information from the calorimeters and muon detectors to select events
at a rate of around 100\unit{kHz} within a time interval of less than 4\mus. The second level, known as the high-level
trigger, consists of a farm of processors running a version of the full event reconstruction software optimized
for fast processing, and reduces the event rate to around 1\unit{kHz} before data storage.
A more detailed description of the CMS detector, together with a definition of the coordinate system used
and the relevant kinematic variables, can be found in Ref.~\cite{Chatrchyan:2008zzk}.

\section{Trigger requirements and event reconstruction}
\label{sec:triggerEventReconstruction}
The data sample used in this analysis was obtained with the logical OR of event triggers that require either missing transverse momentum larger than 100--120\GeV, or a single lepton
with \pt greater than 24--32\GeV, or a single lepton with $\pt>15\GeV$ accompanied by transverse hadronic energy greater than 350--400\GeV, where the exact thresholds depended on the instantaneous luminosity.
The triggers based on missing transverse momentum quantities alone, without a lepton requirement, have high asymptotic efficiency (about 99\%),
but they only reach the efficiency plateau for \ptmiss larger than 250--300\GeV.
The single-lepton triggers are therefore included to ensure high efficiency at lower values of \ptmiss, and they bring the analysis trigger efficiency
up to nearly 100\% for $\ptmiss > 200\GeV$.

The total trigger efficiency has been studied
as a function of the analysis variables \njets, \nb, \MJ, and \mT, defined later in this section,
in the region with $\ptmiss > 200\GeV$.
The efficiency is close to 100\% and is found to be uniform with respect to these analysis
variables over the three years of data taking. The systematic uncertainty in the trigger efficiency
is estimated to be 0.5\%.

Event reconstruction proceeds from particles identified by the particle-flow (PF)
algorithm~\cite{Sirunyan:2017ulk}, which uses information
from the tracker, calorimeters, and muon systems to identify PF candidates as
electrons, muons, charged or neutral hadrons, or photons.
Charged-particle tracks are required to originate from the event primary $\Pp\Pp$ interaction
vertex, defined as the candidate vertex with the largest value of summed physics-object $\pt^2$.
The physics objects used in this calculation are the jets,
clustered using the anti-\kt jet finding algorithm~\cite{Cacciari:2008gp,Cacciari:2011ma} with
the tracks assigned to candidate vertices as inputs, and the associated missing transverse momentum,
taken as the negative vector sum of the \pt of those jets.

Electrons are reconstructed by associating a charged-particle track with electromagnetic calorimeter 
superclusters~\cite{Khachatryan:2015hwa}. The resulting electron candidates  are
required to have $\pt>20\GeV$ and $\abs{\eta}<2.5$, and to satisfy identification
criteria designed to reject light-parton jets, photon conversions, and electrons produced in the decays
of heavy-flavor hadrons. Muons are reconstructed by associating tracks in the muon system with
those found in the silicon tracker~\cite{Chatrchyan:2012xi}. Muon candidates
are required to satisfy $\pt>20\GeV$ and $\abs{\eta}<2.4$.

To preferentially select leptons that originate in the decay of \PW bosons,
and to suppress backgrounds in which the leptons are produced in the decays
of hadrons containing heavy quarks, leptons are required to be isolated from other PF candidates.
Isolation is quantified using an optimized version of the ``mini-isolation'' variable
originally suggested in Ref.~\cite{Rehermann:2010vq}.
The isolation $I_\text{mini}$ is calculated by summing the transverse momentum of the charged hadrons,
neutral hadrons, and photons within $\Delta R\equiv\sqrt{\smash[b]{(\Delta\phi)^2+(\Delta\eta)^2}}<R_0$ of the
lepton momentum vector  \ptvecell, where $\phi$ is the azimuthal angle in radians and
$R_0$ is given by 0.2 for $\pt^{\ell}\leq 50\GeV$, $(10\GeV)/\pt^{\ell}$ for
$50 <\pt^{\ell}< 200\GeV$, and 0.05 for $\pt^{\ell}\geq 200\GeV$.
Electrons (muons) are then required to satisfy $I_\text{mini}/\pt^{\ell}< 0.1\,(0.2)$.

Jets are reconstructed by clustering charged and neutral PF candidates
using the anti-\kt algorithm~\cite{Cacciari:2008gp}
with a distance parameter of $R=0.4$, as implemented in the \FASTJET package~\cite{Cacciari:2011ma}.
Jets are corrected using a \pt- and $\eta$-dependent jet energy calibration~\cite{Chatrchyan:2011ds}, 
and the estimated energy contribution to the jet \pt from pileup~\cite{Cacciari:2007fd} is subtracted.
Jets are then required to satisfy $\pt>30\GeV$ and $\abs{\eta} < 2.4$, as well as jet identification criteria~\cite{Chatrchyan:2011ds}. Finally, jets that have PF constituents matched to an isolated lepton
are removed from the jet collection. The number of
jets satisfying these requirements is a key quantity in the analysis and is denoted
\njets.

Jets are ``tagged'' as originating from the hadronization of \PQb quarks using the
deep combined secondary vertex (DeepCSV) algorithm~\cite{Sirunyan:2017ezt}.
For the medium working point chosen here, the signal efficiency for identifying
\PQb jets with $\pt> 30\GeV$ in \ttbar events is about 68\%. The probability to
misidentify jets in \ttbar events arising from \PQc quarks
is approximately 12\%, while the probability to misidentify jets associated with
light-flavor quarks or gluons as \PQb jets is approximately 1\%. The number of \PQb-tagged jets is
another key quantity in the analysis and is denoted \nb.

The analysis also makes use of large-radius (\largr) jets, denoted generically with the symbol $J$.
These jets are constructed by clustering the standard \smalr ($R=0.4$) jets described above,
as well as isolated leptons, into \largr ($R=1.4$) jets using the anti-\kt algorithm.
Starting the clustering from \smalr jets takes advantage of the corrections that are applied to these jets.
The masses, $m(J_i)$, of the individual \largr jets reflect the \pt spectrum
and multiplicity of the clustered objects, as well as their angular spread.
By summing the masses of all \largr jets in an event, we obtain the variable \MJ, which is central to the analysis
method:
\begin{linenomath}
\begin{equation}
\MJ = \sum_{J_i = {\text{large-}R\text{ jets}}} m(J_i).
\end{equation}
\end{linenomath}
For \ttbar events with a small contribution from ISR, the distribution of \MJ has an
approximate cutoff at $2m_{\PQt}$~\cite{Khachatryan:2016uwr}.
Thus, in the absence of ISR, the requirement $\MJ > 2m_{\PQt}$ is expected to
remove most of the \ttbar background. In contrast, the \MJ distribution for signal events
typically extends to larger values of \MJ because of the presence of more than two top quarks in the decay chain.
However, as discussed in Refs.~\cite{Khachatryan:2016uwr,Sirunyan:2017fsj}, the presence of a significant amount of ISR in a subset of \ttbar background events
generates a tail at large values of \MJ, and understanding this effect is critical for estimating the remaining background in the analysis.

The missing transverse momentum, \ptvecmiss, is defined as
the negative vector sum of the transverse momenta of all PF
candidates and is calibrated taking into account the jet energy corrections.
Dedicated event filters designed to reject instrumental noise are applied to further improve the
correspondence between the reconstructed and the genuine \ptmiss~\cite{CMS-JME-13-004, CMS-PAS-JME-17-001}.

To suppress backgrounds characterized by the presence of a single
\PW boson decaying leptonically, and without any other significant source of \ptvecmiss apart
from the neutrino from this process, we use the quantity \mT, defined as the
transverse mass of the system consisting of the lepton and the
missing transverse momentum vector,
\begin{linenomath}
\begin{equation}
\mT = \sqrt{2\pt^{\ell}\ptmiss[1-\cos(\Delta\phi_{\ell, \ptvecmiss } )]},
\end{equation}
\end{linenomath}
where $\Delta\phi_{\ell,\ptvecmiss}$ is the difference between the
azimuthal angles of \ptvecell and \ptvecmiss.
For both \ttbar events with a single leptonic \PW decay, and for \wjets events with
leptonic \PW boson decay, the \mT distribution peaks strongly below the \PW boson mass.

Although the event selection requires exactly one identified isolated lepton, backgrounds
can still arise from processes in which two leptons are produced but only one
satisfies the identification and isolation criteria. The dominant contribution to this
type of background arises from \ttbar events with two leptonic \PW boson decays,
including \PW decays involving \Pgt leptons, which can themselves decay into hadrons,
electrons, or muons.
To help suppress such dilepton backgrounds, events are vetoed that contain a broader category
of candidates for the second lepton, referred to as veto tracks, which do not
satisfy the stringent lepton identification requirements.
These include two categories of charged-particle tracks:
isolated leptons satisfying looser identification criteria than lepton candidates, as well as a
relaxed momentum requirement, $\pt>10\GeV$, and isolated charged-hadron PF candidates,
which must satisfy $\pt>15\GeV$. For example, isolated charged hadrons can
arise in \Pgt decays.
For either category, the charge of the veto track must be opposite to that of the identified lepton candidate in the event.
To maintain a high selection efficiency for signal events, lepton veto tracks must also satisfy
a requirement on the quantity \mTii~\cite{Lester:mt2,Barr:2003rg},
\ifthenelse{\boolean{cms@external}}{
\begin{linenomath}
\begin{equation}
  \mTii(\ell,v,\ptvecmiss) =
  \min\limits_{\vec{p}_1+\vec{p}_2 = \ptvecmiss} \big[ \max \left \{ \mT(\vec{p}_{\ell}, \vec{p}_1), \mT(\vec{p}_v,\vec{p}_2)\right \} \big],
  \label{eq:MT2}
\end{equation}
\end{linenomath}
}{
\begin{linenomath}
\begin{equation}
\begin{aligned}
  \mTii(\ell,v,\ptvecmiss) = & \\
  \min\limits_{\vec{p}_1+\vec{p}_2 = \ptvecmiss} \big[& \max \left \{ \mT(\vec{p}_{\ell}, \vec{p}_1), \mT(\vec{p}_v,\vec{p}_2)\right \} \big],
  \label{eq:MT2}
\end{aligned}
\end{equation}
\end{linenomath}
}
where $v$ refers to the veto track. The
minimization is taken over all possible pairs of momenta $\vec{p}_1$ and $\vec{p}_2$ that sum to the \ptvecmiss.  For the dominant background, \ttbar, if the lepton, the veto track, and the
missing transverse momentum all result from a pair of leptonically decaying \PW bosons, \mTii is bounded
above by the \PW boson mass. We improve the signal efficiency by requiring
$\mTii<80\GeV$ for loosely identified leptonic tracks and $\mTii<60\GeV$ for hadronic tracks.

Finally, we define $\ST$ as the scalar sum of the transverse momenta of all
the \smalr jets and all leptons passing the selection.

\section{Event selection and analysis regions}
\label{sec:EvtSelAndRegions}
Using the quantities and criteria defined in Section~\ref{sec:triggerEventReconstruction},
events are selected that have exactly one isolated charged lepton (an electron or a muon),
no veto tracks, \mbox{$\MJ>250\GeV$}, \mbox{$\ST>500\GeV$}, \mbox{$\ptmiss>200\GeV$},
and at least seven (six) \smalr jets if $\ptmiss\leq 500\GeV$ ($\ptmiss>500\GeV$).
At least one of the jets must be tagged as originating from a bottom quark.
After this set of requirements, referred to in the following as the {\it baseline selection}, more
than 85\% of the remaining SM background arises from \ttbar production. The contributions from events
with a single top quark or a \PW~boson in association with jets are each about 4--5\%, while the combined
contribution from $\ttbar\PW$ and $\ttbar\PZ$ events is about 2\%.
The background from QCD multijet events after the baseline selection is very small.
Approximately 40\% of signal T1tttt events are selected with the single-lepton requirement only.

\begin{figure*}[tbp!]
\centering
  \includegraphics[width=\textwidth]{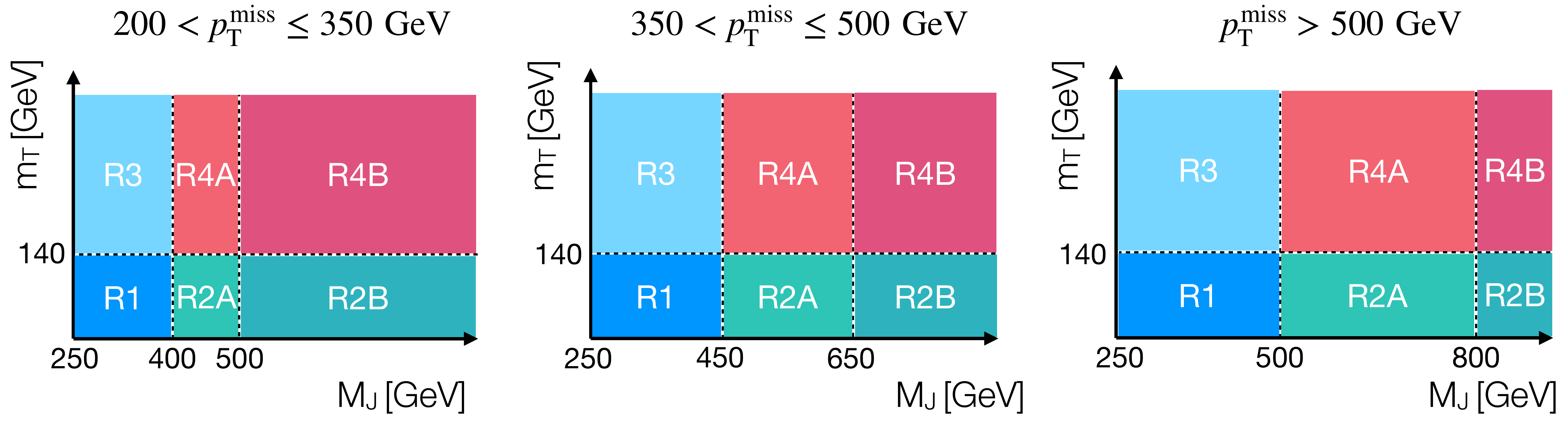}
\caption{Analysis regions defined for each bin in \ptmiss. For the signal models considered here, the
regions R1, R2A, R2B, and R3 are dominated by background, while R4A and R4B would have a
significant signal contribution. In the combined fit performed to the event yields observed in these regions, signal contributions
are allowed in the background-dominated regions. The R2A, R2B, R4A, and R4B regions are further divided into bins of
\njets and \nb, as discussed in the text. }
\label{fig:abcd}
\end{figure*}

To improve the sensitivity to the signal and to provide a method for the background estimation,
the events satisfying the baseline selection are divided into a set of
signal and control regions in the \MJ-\mT plane and in bins of
\ptmiss, \njets, and \nb. In each of the three \ptmiss regions, $200<\ptmiss\leq 350\GeV$,
$350<\ptmiss\leq 500\GeV$, and $\ptmiss > 500\GeV$, the \MJ-\mT plane is divided into
six regions, referred to as R1, R2A, R2B, R3, R4A, and R4B, as shown in
Fig.~\ref{fig:abcd}. The signal regions are R4A and R4B, while R1, R2A, R2B, and R3 serve as
control regions. Potential signal contamination in the control regions is taken
into account using a fit method described in Section~\ref{sec:backgroundEstimation}.
Regions denoted with the letter A are referred to as low \MJ, while regions denoted with
the letter B are referred to as high \MJ.
The control regions R1, R2A, and R3 are used
to estimate the background in signal region R4A, while the control regions R1, R2B, and R3
are used to estimate the background in signal region R4B. (In discussions
where the distinction between R2A and R2B, or between R4A and R4B, is irrelevant, we refer
to these regions generically as R2 and R4.)
As seen in Fig.~\ref{fig:abcd}, for each of the three regions in \ptmiss,
the \MJ ranges for R2A and R4A (low \MJ) and for R2B and R4B (high \MJ) are

\begin{itemize}
\item $200<\ptmiss\leq 350\GeV$: $400<\MJ\leq 500\GeV$ (low \MJ) and $\MJ> 500\GeV$ (high \MJ)
\item $350<\ptmiss\leq 500\GeV$: $450<\MJ\leq 650\GeV$ (low \MJ) and $\MJ> 650\GeV$ (high \MJ)
\item $\ptmiss> 500\GeV$: $500<\MJ\leq 800\GeV$ (low \MJ) and $\MJ> 800\GeV$ (high \MJ).
\end{itemize}

The use of six regions in the \MJ-\mT plane (in each bin of \ptmiss) is an improvement over the original method
used in Refs.~\cite{Khachatryan:2016uwr,Sirunyan:2017fsj}, where only four regions were used: R1, R2 (combining
R2A and R2B), R3, and R4 (combining R4A and R4B). The larger event yields
in the full Run 2 data sample allow for this additional division of the \MJ-\mT plane.
By separating each of the original ``high'' \MJ regions into two bins, we are able to obtain additional sensitivity to
SUSY models with large mass splittings, which tend to populate the highest
\MJ regions with a significant number of events.
In addition, the values of \MJ corresponding to the boundaries between these regions increase with \ptmiss,
improving the expected precision in the background prediction.

Regions R2A, R2B, R4A, and R4B are further subdivided into bins of \njets and \nb to increase sensitivity to the signal:
\begin{itemize}
\item two \njets bins: $\njets=7$ ($6\leq\njets\leq7$) for $\ptmiss\le 500\GeV$ ($\ptmiss>500\GeV$) and $\njets\geq8$
\item three \nb bins: $\nb=1$, $\nb=2$, and $\nb\geq3$.
\end{itemize}
The total number of signal regions is therefore $3 (\ptmiss) \times 2(\MJ) \times 2(\njets) \times 3(\nb) = 36$.
Given that the main background processes have two or fewer \PQb quarks, the total SM contribution to the
$\nb\geq3$ bins is very small and is driven by the \PQb jet mistag rate.  Signal
events in the T1tttt model are expected to populate primarily the bins with $\nb\geq2$,
while bins with $\nb=1$ mainly serve to test the method in a background dominated region.

Because of the common use of R1 and R3 in the background
estimations for R4A and R4B, as well as the integration over \njets and \nb in the R1 and R3 regions,
there are statistical correlations between the background predictions,
which are taken into account in the fitting methodology (Section~\ref{sec:backgroundEstimation}).

\section{Background estimation method}
\label{sec:backgroundEstimation}
\begin{figure}[tb!]
\centering
\includegraphics[width=\cmsFigWidth]{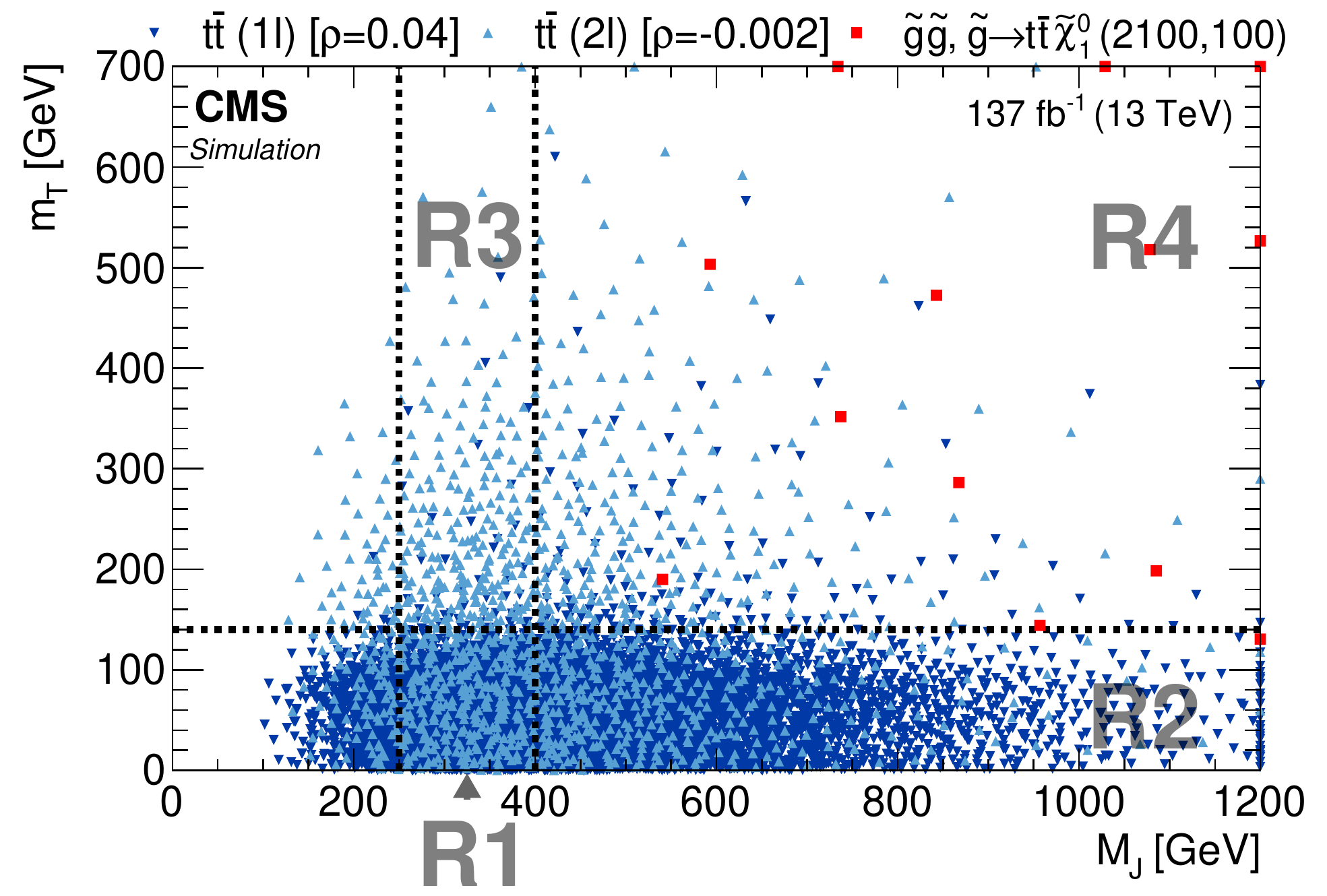}
\caption{Distribution of simulated single-lepton \ttbar events (dark-blue inverted triangles) and dilepton \ttbar events
(light-blue triangles) in the \MJ-\mT
plane after applying the baseline selection and requiring at least two \PQb jets. A representative
random sample of T1tttt events with $\mGlu=2100\GeV$ and $\mLSP=100\GeV$ is also shown for comparison (red squares).
Each marker represents one expected event in the full data sample. Overflow
events are placed on the edges of the plot. The values of the
correlation coefficients $\rho$ for each of the background processes are given in the legend. Region R4,
which is further split into smaller bins R4A and R4B, is the nominal signal region,
while R1, R2, and R3 serve as control regions. This plot is only illustrative, because
the boundary between R1 and R2, as well as between R3 and R4, is \ptmiss-dependent. The line shown at 400\GeV corresponds to the value used for the lowest \ptmiss bin.
Additional sensitivity is obtained by binning the events in \ptmiss, \njets, and \nb.}
\label{fig:scat_baseline}
\end{figure}

The method for estimating the background yields in each of the signal bins takes advantage of the fact that the \MJ
and \mT distributions of background events with a significant amount of ISR are largely uncorrelated
and that there are background-dominated control samples that
can be used to test the method and establish systematic uncertainties.
Figure~\ref{fig:scat_baseline} shows the two-dimensional distribution of simulated \ttbar events in the variables
\MJ and \mT, with single-lepton and dilepton events shown with separate symbols.
The three background-dominated regions (R1, R2, and R3)
and the signal region (R4) are indicated. (For simplicity, the separate A and B regions for R2 and R4 are not shown
in this figure.) The low-\mT region, \mbox{$\mT\leq140\GeV$},
is dominated by \ttbar single-lepton events, and the rapid falloff in the number of such events
as \mT increases is apparent. In contrast, the high-\mT region, \mbox{$\mT> 140\GeV$}, is dominated by \ttbar dilepton
events. As discussed in Refs.~\cite{Khachatryan:2016uwr,Sirunyan:2017fsj}, the
\MJ distributions for the events in these two regions become nearly identical in the
presence of significant ISR, which is enforced by the jet multiplicity
requirements. This behavior allows us to measure the shape of the \MJ distribution
at low \mT with good statistical precision and then use it to obtain
a background prediction in the high-\mT region by normalizing it to the event yield in R3.

To estimate the background contribution in each of the signal bins, a modified version of an ``ABCD'' method
is used. Here, the symbols A, B, C, and D refer to four regions in a two-dimensional space in the data, where
one of the regions is dominated by signal and the other three are dominated by backgrounds.
In a standard ABCD method, the background rate ($\mu^\text{bkg}_\text{region}$) in the signal region (in this case, either R4A or R4B) is estimated from the yields ($N_\text{region}$) in
three control regions using
\begin{linenomath}
\begin{equation}
\label{eq:bkgest:abcd:nokappa}
  \begin{aligned}
\mu^\text{bkg}_\text{R4A} &= \frac{N_\text{R2A} N_\text{R3}}{N_\text{R1}},\\
\mu^\text{bkg}_\text{R4B} &= \frac{N_\text{R2B} N_\text{R3}}{N_\text{R1}},
  \end{aligned}
\end{equation}
\end{linenomath}
where the labels of the regions correspond to those shown in Fig.~\ref{fig:abcd}.
The background prediction is unbiased in the limit that the
two variables that define the plane (in this case, \MJ and \mT) are uncorrelated.
The effect of any residual correlation can be taken into account
by multiplying these background predictions by correction
factors $\kappa_A$ and $\kappa_B$,
\begin{linenomath}
\begin{equation}
\label{eq:bkgest:abcd}
  \begin{aligned}
\mu^\text{bkg}_\text{R4A} &= \kappa_A \left(\frac{N_\text{R2A}N_\text{R3}}{N_\text{R1}}\right),\\
\mu^\text{bkg}_\text{R4B} &= \kappa_B \left(\frac{N_\text{R2B}N_\text{R3}}{N_\text{R1}}\right),
  \end{aligned}
\end{equation}
\end{linenomath}
which are double ratios obtained from simulated event samples:
\begin{linenomath}
\begin{equation}
\label{eq:bkgext:kappa}
  \begin{aligned}
\kappa_A &= \frac{N_\text{R4A}^\text{MC,bkg}/N_\text{R3}^\text{MC,bkg}}{N_\text{R2A}^\text{MC,bkg}/N_\text{R1}^\text{MC,bkg}},\\
\kappa_B &= \frac{N_\text{R4B}^\text{MC,bkg}/N_\text{R3}^\text{MC,bkg}}{N_\text{R2B}^\text{MC,bkg}/N_\text{R1}^\text{MC,bkg}}.
  \end{aligned}
\end{equation}
\end{linenomath}
When the two ABCD variables are uncorrelated, or nearly so, the $\kappa$ factors are close to unity.
This procedure ignores potential signal contamination in the control regions, which is
accounted for by incorporating the methods described above
into a fit that includes both signal and background components.

In principle, this calculation to estimate the background can be performed for each of the 36 signal
bins by applying this procedure in 36 independent ABCD planes. However, such an approach would incur large statistical
uncertainties in some bins due to the small number of events in R3 regions. This problem is especially important in bins with a
large number of jets, where the \MJ distribution shifts to higher values and the number of background events
expected in R4 can even exceed the background in R3.

To alleviate this problem, we exploit the fact that, after the baseline selection, the background is dominated
by a single source (\ttbar events), and the shapes of the \njets distributions are nearly identical
for the single-lepton and dilepton components, because of the large amount of ISR.
As a result, the \mT distribution is approximately independent of \njets and \nb.
More specifically, we find that for \MJ values corresponding to the R1 and R3 regions,
the ratios of high-\mT to low-\mT event yields do not vary substantially between events with seven or more jets, and across \nb within these \njets bins. We exploit this result by integrating the event yields in the low-\MJ regions (R1 and R3) over
the \njets and \nb bins for each \ptmiss bin. This procedure increases the statistical power of the ABCD method
but also introduces a correlation among the predictions from Eq.~\eqref{eq:bkgest:abcd}
for the \njets and \nb bins associated with a given \ptmiss bin.

\begin{figure*}[tbp!]
  \includegraphics[width=0.49\textwidth]{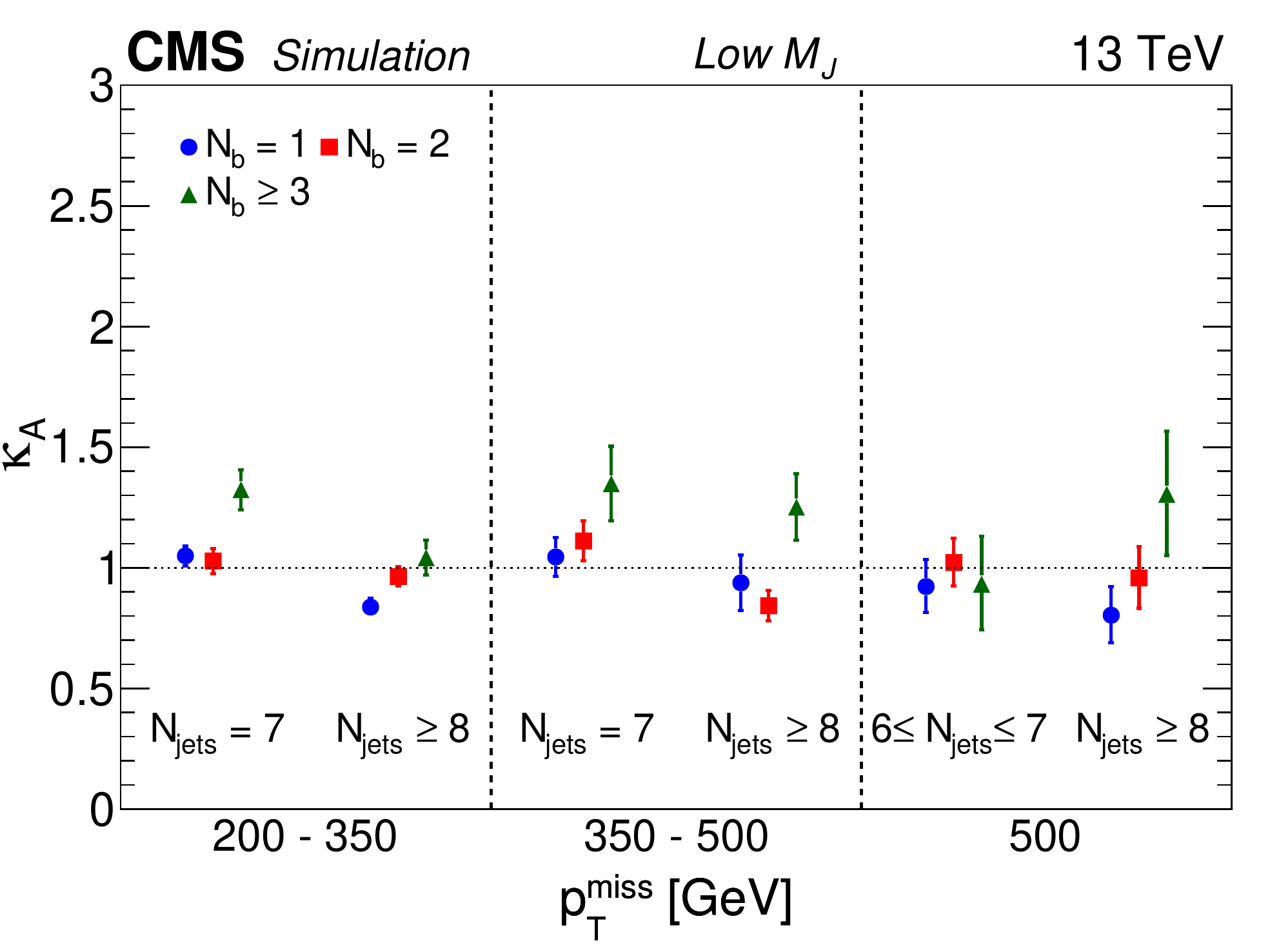}
  \includegraphics[width=0.49\textwidth]{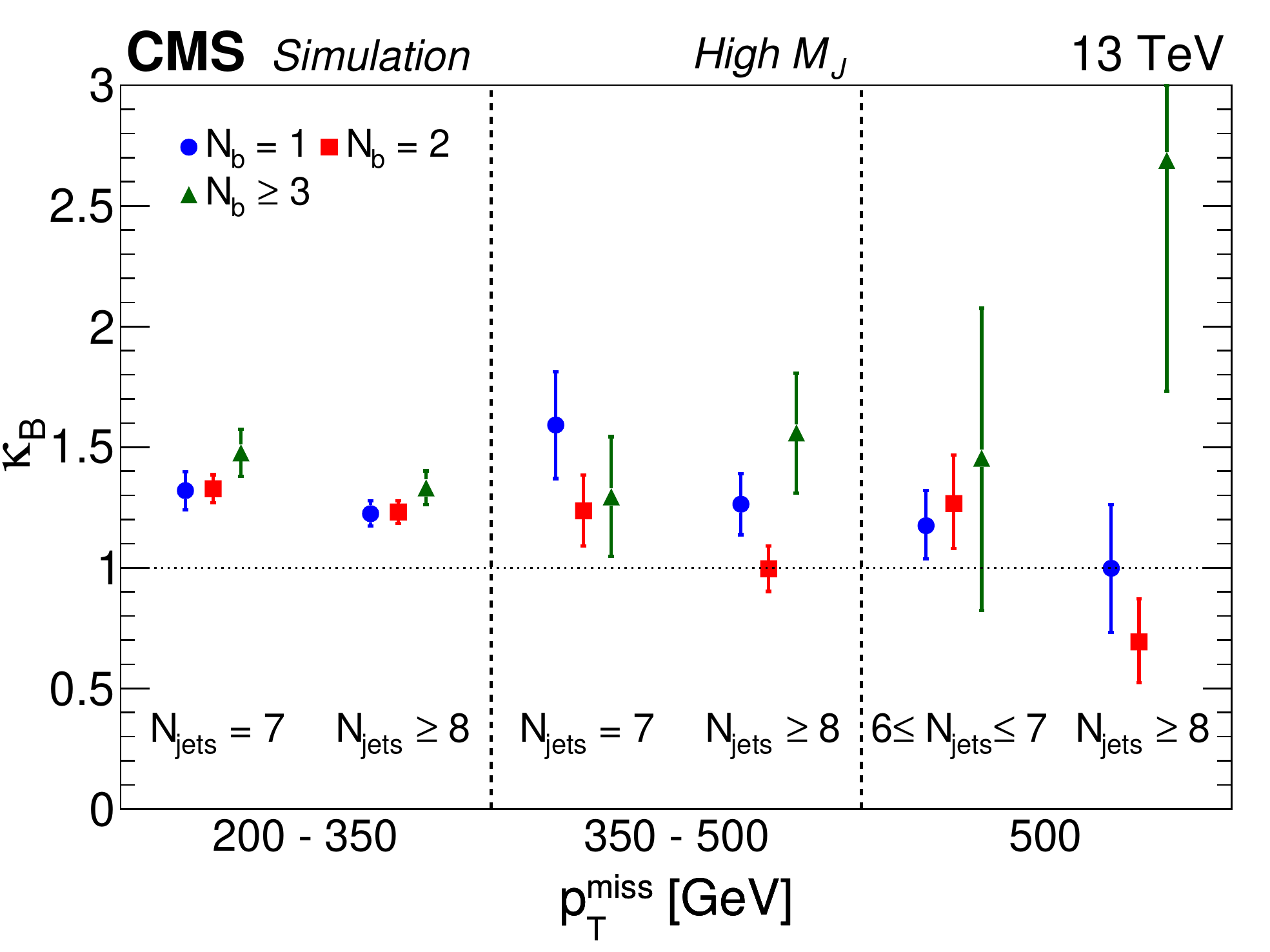}
  \centering \caption{Values of the double-ratio $\kappa$ for each of the 18 signal bins of the \lowmj ABCD planes, \ie, R1-R2A-R3-R4A (left),
  and the 18 signal bins of the \highmj ABCD planes, \ie, R1-R2B-R3-R4B (right), calculated
    using the simulated SM background. The $\kappa$ factors are close to unity,
    indicating a small correlation between \MJ and \mT. The uncertainties shown are statistical only.
    \label{fig:bkgest:kappa}}
\end{figure*}

Figure~\ref{fig:bkgest:kappa} shows the values of the $\kappa$ factors obtained from simulation
(computed after integrating over \njets and \nb in R1 and R3 only)
for the 18 signal bins of the \lowmj ABCD planes, \ie, R1-R2A-R3-R4A (left plot),
and the 18 signal bins of the \highmj ABCD planes, \ie, R1-R2B-R3-R4B (right plot).
These values are close to unity for the \lowmj regions and are slightly above unity for the \highmj regions. 
The deviation from unity is due to the presence of mismeasured jets in single-lepton \ttbar events, which produces a correlation between \mT and \MJ. The additional \ptmiss arising from the jet mismeasurement allows these events to migrate from the low-\mT to the high-\mT region.  Since the mismeasured \ptmiss is correlated with hadronic activity, these events typically also have larger \MJ values relative to well-reconstructed events. Consequently, their presence at high \mT results in a difference between the shapes of the \MJ distributions for low-\mT and high-\mT events and thus results in a $\kappa$ value larger than unity. In addition to the statistical uncertainties shown in Fig.~\ref{fig:bkgest:kappa}, systematic uncertainties are obtained from studies of the modeling of the $\kappa$ values in dedicated data control samples, including both a sample with high purity of dilepton \ttbar events as well as a sample enriched in mismeasured single-lepton \ttbar events, as discussed in Section~\ref{sec:sysUncert}.

The method described above is implemented with a maximum likelihood fit to the event yields observed in data using a
likelihood function that incorporates both the statistical and systematic uncertainties in $\kappa_A$ and $\kappa_B$.
The fit also takes into account the correlations associated with the common R1 and R3 yields that arise
from the integration over \njets and \nb, and it accounts
for the signal contamination in the control regions.

The signal strength is the only parameter that enters the likelihood in a way that extends across \ptmiss bins.
We can therefore define the correlation model within each \ptmiss bin and then take the product over
\ptmiss bins to construct the full likelihood function. Let $\mu_{i,j,k}^\text{bkg}$ be the estimated
(Poisson) mean background in each region, where $i$ indicates the \ptmiss bin, $j\in S$ with $S\equiv$
\{R1, R2A, R2B, R3, R4A, R4B\}, and $k$ runs over the six \njets-\nb bins.
Then, in a given \ptmiss bin, the 26 background rates (one each for R1 and R3 and six each for R2A, R2B, R4A, and R4B)
can be expressed in terms
of 14 floating fit parameters $\theta$ (one each for R1 and R3 and six each for R2A and R2B)
and the 12 correction factors $\kappa$ ($\kappa_A$ and $\kappa_B$ for each of the six \njets and \nb bins for a fixed \ptmiss bin) as
\begin{linenomath}
\begin{equation}
\label{eq:rateParams}
\begin{aligned}
  \mu_{\text{R1}}^{\text{bkg}}=\theta_{\text{R1}},
  &&\mu_{\text{R2A},k}^{\text{bkg}}=&\theta_{\text{R2A},k}, \\
  &&\mu_{\text{R2B},k}^{\text{bkg}}=&\theta_{\text{R2B},k},\\
  \mu_{\text{R3}}^{\text{bkg}}=\theta_{\text{R3}},
  &&\mu_{\text{R4A},k}^{\text{bkg}}=&\kappa_{\text{A},k} \, \theta_{\text{R2A},k} \, (\theta_{\text{R3}} / \theta_{\text{R1}}),\\
  &&\mu_{\text{R4B},k}^{\text{bkg}}=&\kappa_{\text{B},k} \, \theta_{\text{R2B},k} \, (\theta_{\text{R3}} / \theta_{\text{R1}}).
\end{aligned}
\end{equation}
\end{linenomath}
Here, the $i$ index for the three \ptmiss bins is suppressed, because it applies identically to all parameters in the equations.
In addition, the $k$ index over the \njets and \nb bins is omitted for terms that are integrated over these quantities, \ie, for the parameters for the R1 and R3 regions.
The quantity $\theta_{\text{R3}} / \theta_{\text{R1}}$ is simply the ratio between the background event rates in regions R3 and R1, summed over \njets and \nb. To obtain the prediction for the mean background,
this ratio is then multiplied by the appropriate rate
$\theta_{\text{R2A},k}$ or $\theta_{\text{R2B},k}$ and then corrected with the appropriate value $\kappa_A$ or $\kappa_B$ for the
given bin in \njets and \nb.

Defining $N^\text{data}_{i,j,k}$ as the observed data yield in each region and bin,
$\mu_{i,j,k}^{\text{MC,sig}}$ as the corresponding expected signal rate, and $r$ as the
parameter quantifying the signal strength relative to the expected yield across all
analysis regions, we can write the likelihood function as
\begin{linenomath}
\begin{equation}
\label{eq:bkgest:likelihood}
\begin{aligned}
  \mathcal{L}&=\prod_{i}^{\ptmiss\text{bins}}\mathcal{L}_{\text{ABCD},i}^\text{data}\, \mathcal{L}^\text{MC}_{\text{sig},i},\\
  \mathcal{L}_{\text{ABCD},i}^\text{data} &= \prod_{j\in S}\prod_{k=1}^{N_\text{bins}(j)}
  \text{Poisson}(N^\text{data}_{i,j,k}|\mu_{i,j,k}^{\text{bkg}}+r\,\mu_{i,j,k}^{\text{MC,sig}}),\\
  \mathcal{L}^{\text{MC}}_{\text{sig},i} &= \prod_{j\in S}\prod_{k=1}^{N_\text{bins}(j)}\text{Poisson}(N_{i,j,k}^{\text{MC,sig}}|\mu_{i,j,k}^{\text{MC,sig}}).
\end{aligned}
\end{equation}
\end{linenomath}
Given the integration over \njets and \nb in R1 and R3, $N_\text{bins}(\text{R}1)=N_\text{bins}(\text{R}3)=1$, while
$N_\text{bins}(\text{R2A})=N_\text{bins}(\text{R2B})=N_\text{bins}(\text{R4A})=N_\text{bins}(\text{R4B})=6$.

In Eq.~\eqref{eq:bkgest:likelihood}, the $\mathcal{L}_{\text{ABCD},i}^\text{data}$ terms account for the statistical
uncertainty in the observed data yield in each bin, and
the $\mathcal{L}^{\text{MC}}_{\text{sig},i}$ terms account for the uncertainty in the signal shape, due to the finite size of the MC samples. The statistical uncertainties in the $\kappa$ factors due to the finite size of the simulated background event samples are implemented as Gaussian constraints.
The signal systematic uncertainties are incorporated in the likelihood function as log-normal constraints with a nuisance
parameter for each uncorrelated source of uncertainty. These terms are not explicitly shown in the likelihood function above for simplicity.

The likelihood function defined in Eq.~\eqref{eq:bkgest:likelihood} is employed in two
separate types of fits that provide complementary but compatible background estimates based on an ABCD model.
The ``R1--R3 fit'' is used to test the agreement between the observed event yields (R4) and the predictions (based on R1, R2, and R3 event
yields) under the null (\ie, the background-only) hypothesis. In this approach, we exclude the observations in the signal
regions in the likelihood and fix the signal strength $r$ to 0. This procedure involves as many unknowns as constraints.
As a result, the estimated background rates in regions R1, R2, and R3 become simply the observed values in those bins, and we
obtain predictions for the signal regions that do not depend on the observed $N^\text{data}_{\text{R}4}$.  The
R1--R3 fit thus corresponds to the standard ABCD method with $\kappa$ corrections, and the likelihood machinery becomes just a convenient
way to solve the system of equations and propagate the various uncertainties.

In contrast, the ``R1--R4 fit'' also makes use of the observations in the signal regions, and it can therefore
provide an estimate of the signal strength $r,$ while also allowing for signal events to populate the control
regions. We also use the R1--R4 fit with the constraint $r=0$ to assess the agreement between the data and the background predictions in the
null hypothesis.

\section{Background systematic uncertainties}
\label{sec:sysUncert}
The background estimation procedure described in Section~\ref{sec:backgroundEstimation}
relies on the approximate independence of the kinematic variables \MJ and \mT, as well as on
the ability of the simulation to correctly model any residual correlation, which would manifest as a departure
of $\kappa$ from unity. The approximate independence of \MJ and \mT
is a consequence of two key features of the data, namely, that the high-\mT sample
is composed primarily of dilepton \ttbar events and that the \MJ spectra of \ttbar events
with one and two leptons become highly similar in the presence of ISR jets. A residual correlation of \MJ and \mT can arise either from (i) contributions to the overall \MJ shape from backgrounds other than single-lepton \ttbar at low \mT and dilepton \ttbar at high \mT or from (ii) subleading kinematic effects that result in the gradual divergence of the single-lepton and dilepton \MJ shapes as a function of the analysis binning variables. As an example of (i), simulation studies show that the deviation of $\kappa$ from unity for the high-\MJ ABCD planes, most pronounced at low \ptmiss, can arise from mismeasured single-lepton \ttbar events that populate the high-\mT region. A classification and study of such mechanisms was presented in Ref.~\cite{Khachatryan:2016uwr}. Based on this understanding, the systematic uncertainties in the background estimate are obtained by quantifying the ability of the simulation to predict the behavior of $\kappa$ in control samples in data with varying background composition and as a function of the analysis binning variables.

\subsection{Control sample strategy}
\label{ssec:strategy}

Two control samples are used to assess the ability of the simulation to reproduce the behavior of $\kappa$ in the data:
a $2\ell$ sample composed of events with two reconstructed leptons and a $1\ell$, 5--6 jet sample composed of events with a single reconstructed lepton and
either five or six jets.

Because it is composed primarily of \ttbar dilepton events, the $2\ell$ control sample allows us to assess the validity of the main
assertion of the analysis, namely that the shapes of the \MJ distributions for $1\ell$ and $2\ell$ \ttbar events
approximately converge at high jet multiplicities. The MC predictions for $\kappa$ are tested independently as a function of \njets and \ptmiss using this control sample,
because simulation studies show no significant correlation in the $\kappa$ behavior as a function of these two variables.
The dilepton control sample is not used to probe the modeling of $\kappa$ as a function of $\nb$, which is instead studied in the 5--6 jet control sample described below. Events in the dilepton control sample with $\nb\geq2$ are excluded to avoid potential signal contamination.

Except for the case of the dilepton \ttbar process, it is not possible to find useful control samples where a
particular background category dominates. As a consequence, we cannot completely factorize the uncertainty
in $\kappa$ arising from mismodeling of the background composition and from mismodeling the \mT-\MJ
correlation for a particular background. However, we are able to define
a control sample in which the background composition and kinematic characteristics are very
similar to those in the signal regions, but in which the expected signal contribution is too small to significantly affect the
data vs.~simulation comparison. The single-lepton, 5--6 jet sample satisfies these
requirements. Both the $\kappa$ values for individual background categories and the composition
of background processes are very similar to those for events with $\njets\geq7$. We therefore
use this control sample to quantify mismodeling of $\kappa$ arising either from detector mismeasurement effects
(which can result in a larger fraction of single-lepton \ttbar events at high \mT),
or from mismodeling of the background composition.
An \nb-dependent uncertainty is derived from the lowest \ptmiss region (which is binned in \nb).
Based on studies in simulation, any \nb dependence is not correlated with \ptmiss within the statistical
precision of the sample, and therefore the uncertainties derived in the low-\ptmiss region can be used for all \ptmiss bins.
Since the low-\ptmiss bin has the highest contribution from events with \ptmiss mismeasurement, this uncertainty
also provides an estimate of the uncertainty in the modeling of $\kappa$ in the presence of mismeasurement that is valid over the full \ptmiss range.
We have verified in simulation that artificially increasing the fraction of mismeasured events has a consistent effect across the bins
in the single-lepton, 5--6 jet control sample and the corresponding signal bins, so this effect would be detected in a study of this control sample.

\subsection{Dilepton control sample results}
\label{ssec:cr2lveto}

We construct an alternate ABCD plane in which the high-\mT regions R3
and R4A/B are replaced with regions D3 and D4A/B, which are defined as having either two reconstructed leptons
or one lepton and one isolated track. The new regions D3 and D4A/B are constructed to mimic the
selection for R3 and R4A/B, respectively. For the events with two
reconstructed leptons in D3 and D4A/B, the selection is modified as follows: the \njets bin
boundaries are lowered by 1 to keep the number of large-$R$ jet constituents the same as in the
single-lepton samples; the \mT requirement is not applied; and events with both $\nb=0, 1$ are included
to increase the size of the event sample. The lepton-plus-track
 events in D3 and D4A/B are required to pass the same selection as those in R3 and R4 except for the track veto.
With these requirements, the sample is estimated from simulation to consist of between 75--85\% \ttbar dilepton
events, depending on the \ptmiss and \njets bin.

\begin{figure*}[bth!]
  \centering
  \includegraphics[width=0.49\textwidth]{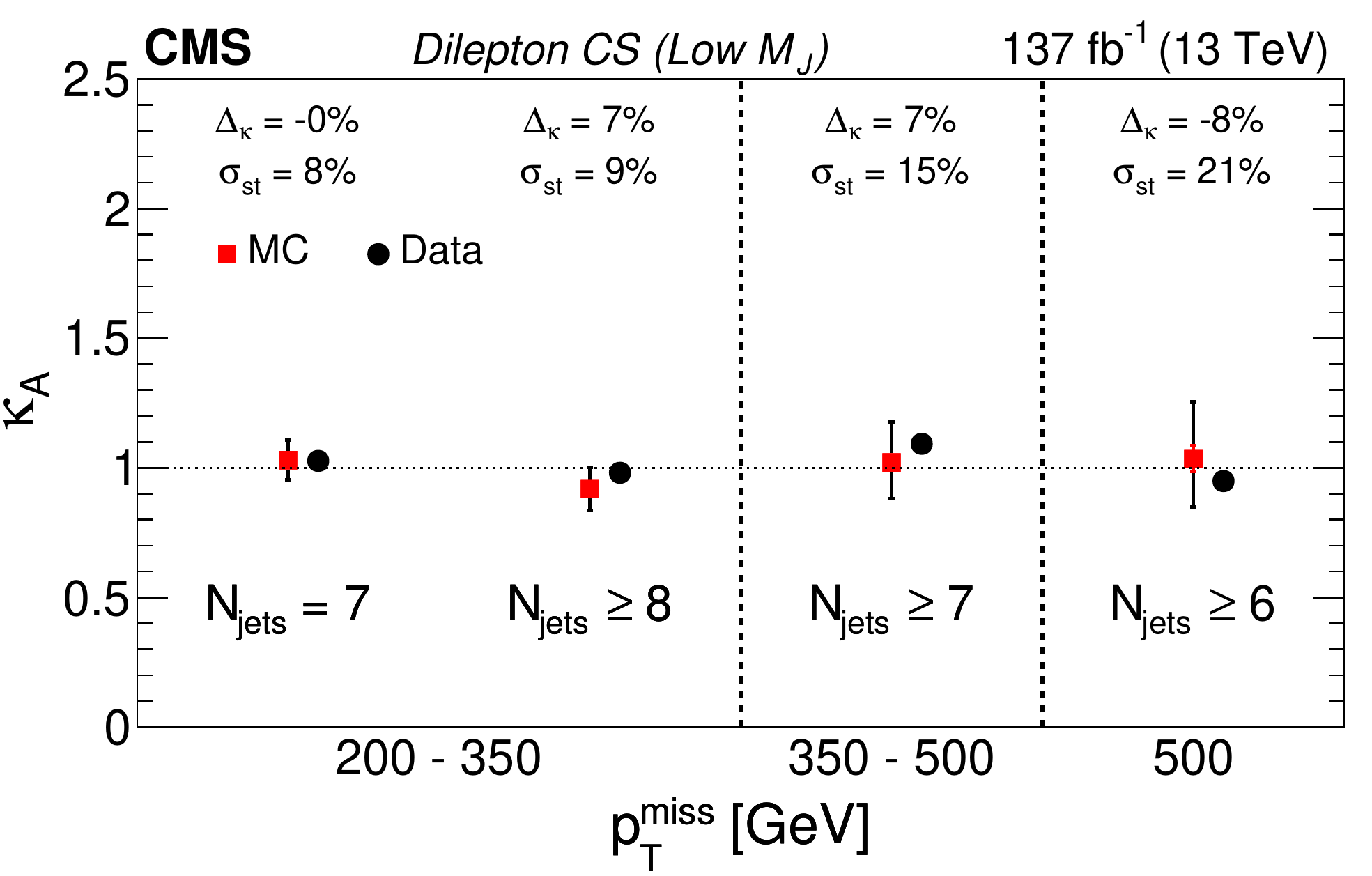}
  \includegraphics[width=0.49\textwidth]{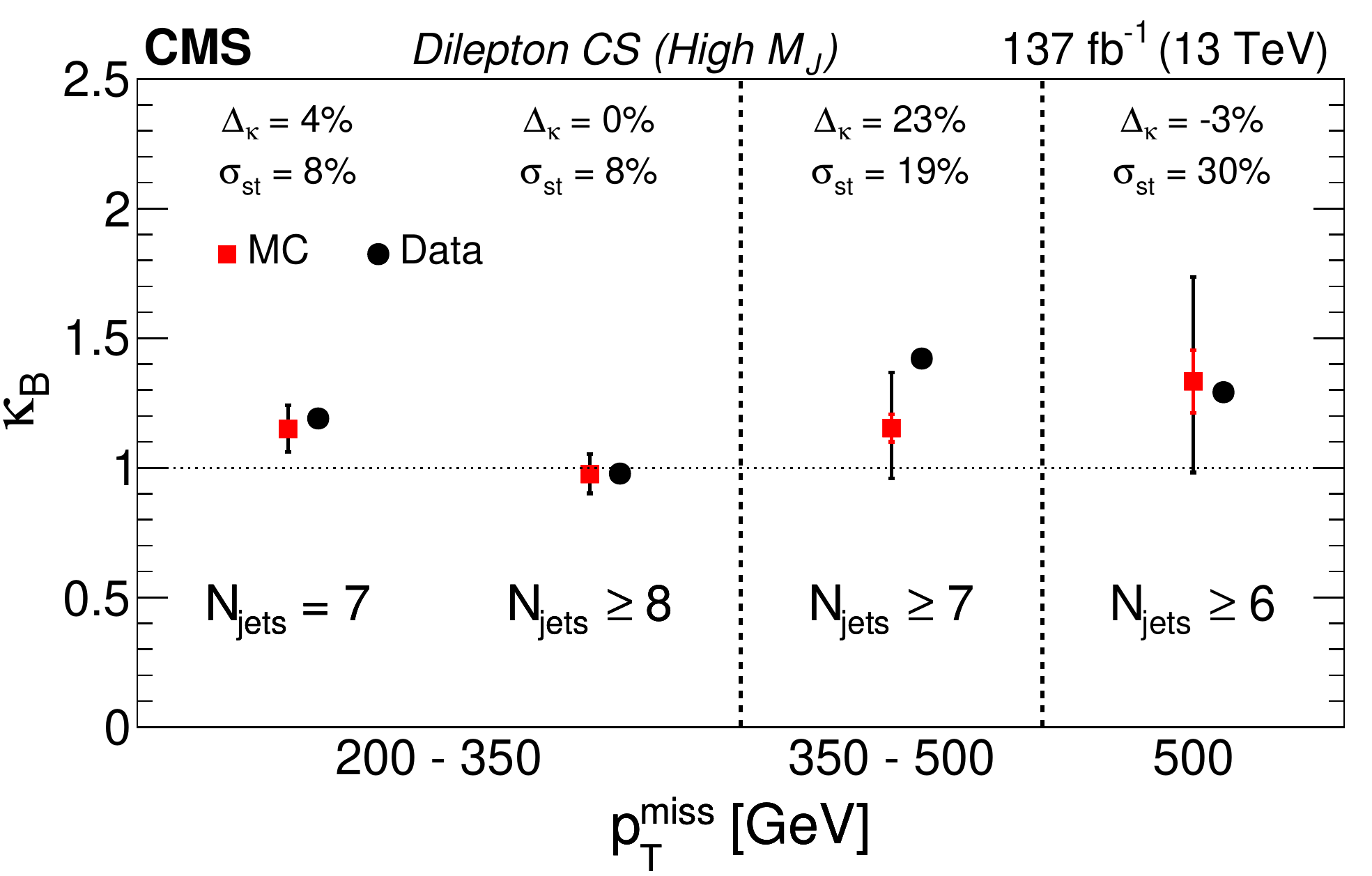}
  \caption{Dilepton control sample (CS): validation of the $\kappa$ factor values found in simulation vs.~data
  for low \MJ (left) and high \MJ (right). The data and simulation are shown as black and red points,
  respectively. No statistical uncertainties are plotted for the data points,
  but instead, the expected statistical uncertainty for the data points, summed in quadrature with
  the statistical uncertainty of the simulated samples, is given by the error bar on the red points and is quoted as $\sigma_{\mathrm{st}}$.
  The red portion of the error bar on the red points indicates the contribution from the simulated samples.
  The quoted values of $\Delta_{\kappa}$ are defined as the relative difference between the $\kappa$ values found in simulation and in data.
   }
  \label{fig:syst:cr_data_dileptons}
\end{figure*}

Using the dilepton control sample, we compute the values of $\kappa$ in both simulation and data
in the two \njets bins at low \ptmiss, and integrated over \njets in the intermediate- and high-\ptmiss bins.
Figure~\ref{fig:syst:cr_data_dileptons} compares the $\kappa$ values obtained from simulation with those observed in
data in the dilepton control sample. We observe that these values are consistent within the total statistical
uncertainties, and we therefore assign the statistical uncertainty in this comparison as the systematic uncertainty in
$\kappa$ as follows. We take the uncertainty associated with the \njets dependence of $\kappa$ from the lowest \ptmiss bin, specifically, 8 (8)\% for low \njets and 9 (8)\% for high \njets at low \MJ (high \MJ), and use these values in the intermediate- and high-\ptmiss bins as well. This procedure is based on the observation that in simulated event samples the dependence of $\kappa$ on \njets is consistent across \ptmiss bins. This uncertainty also accounts for potential mismodeling of $\kappa$ at low \ptmiss. Then, to account for additional possible sources of mismodeling of
$\kappa$  as a function of \ptmiss, we assign an uncertainty based on the comparison
between simulation and data at intermediate- and high-\ptmiss values for low \MJ (high \MJ) as 15 (19)\% and 21 (30)\%, respectively. These uncertainties contribute to the total uncertainty for each signal region,
as summarized in Section~\ref{ssec:sys_summary}.

\subsection{Single-lepton, 5--6 jet control sample results}
\label{ssec:cr56j}

The single-lepton, 5--6 jet control sample (referred to simply as the 5--6 jet control sample)
is constructed in a manner identical to the signal
regions, except for the \njets requirement. The $\kappa$ values are studied in the low-
and intermediate-\ptmiss bins, while the high-\ptmiss bin is not considered because of potential signal contamination
(6-jet events are in fact part of the signal regions at high \ptmiss).

\begin{figure*}[bth!]
  \centering
  \includegraphics[width=0.49\textwidth]{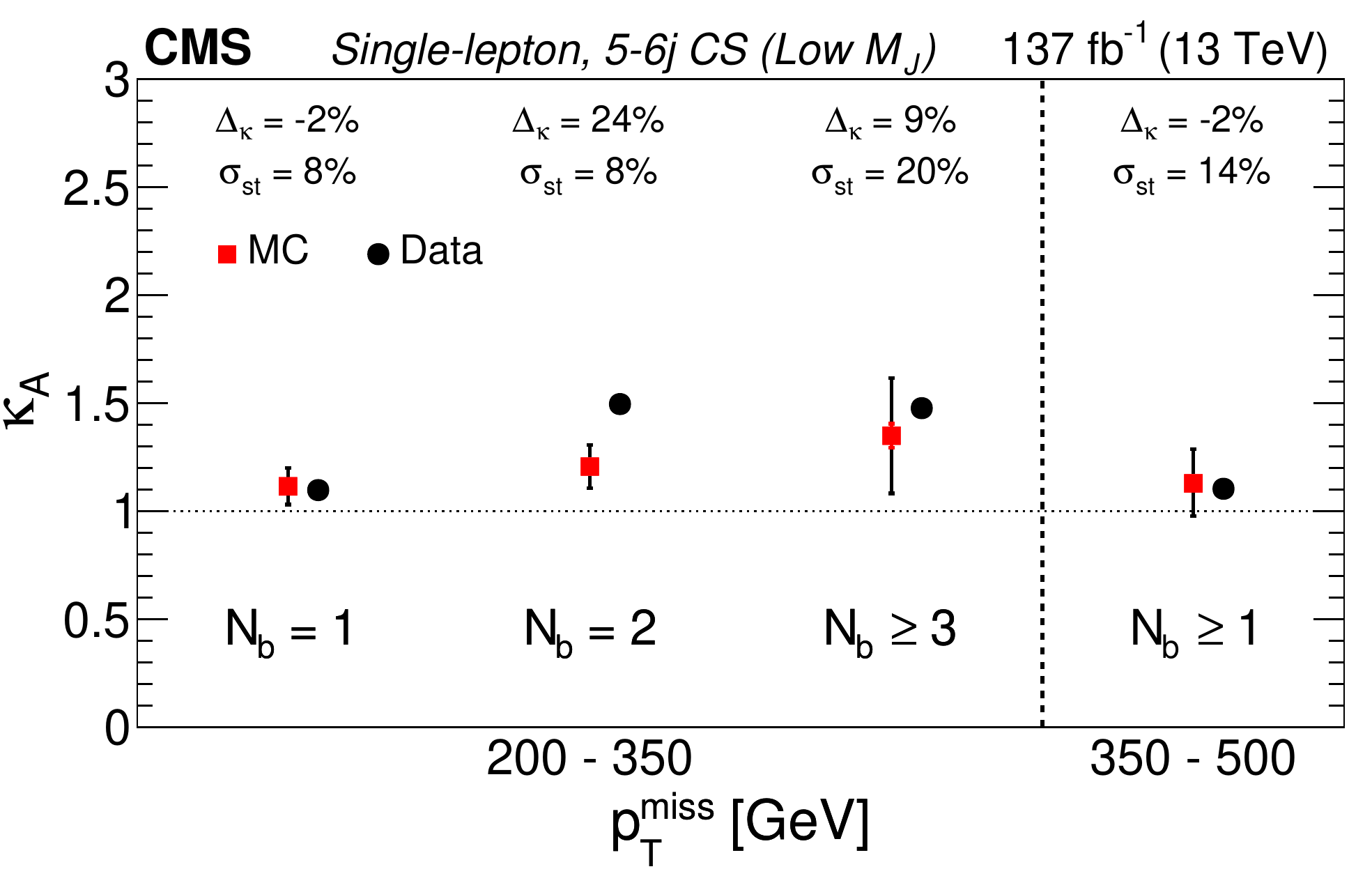}
  \includegraphics[width=0.49\textwidth]{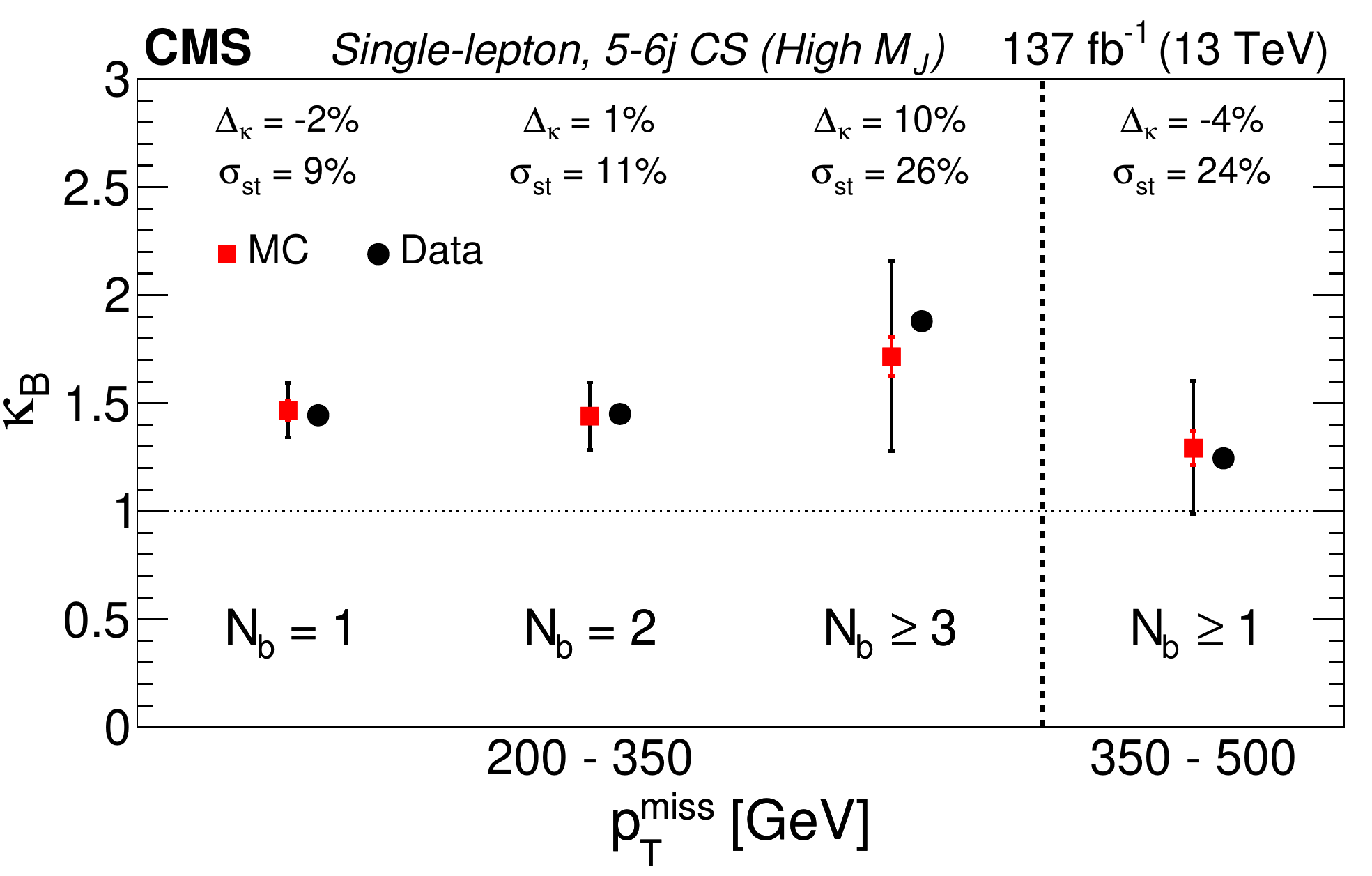}
  \caption{Single-lepton 5--6 jet CS: validation of the $\kappa$ factor values found in simulation vs.~data for low \MJ (left) and high \MJ (right). The data and simulation are shown as black and red points, respectively. The expected uncertainty in the data, summed in quadrature with the statistical uncertainty of the simulated samples, is given by the error bar on the red points ($\sigma_{\mathrm{st}}$). The red portion of the error bar indicates the contribution from the simulated samples. The values of $\Delta_{\kappa}$ are the relative difference between the $\kappa$ values found in simulation and in data.}
\label{fig:syst:cr_data_56jets}
\end{figure*}

The $\kappa$ measurement is performed in the three \nb bins at low \ptmiss
and is also performed in the intermediate \ptmiss bin, integrating over \nb. Figure~\ref{fig:syst:cr_data_56jets}
compares the $\kappa$ values obtained from simulation with those measured in
the data. We find consistency between the simulation and the data except for a $3\sigma$ deviation in the 2 \PQb-jet bin.
Closer examination of distributions contributing to this $\kappa$ value shows a higher yield in the region equivalent to R4A
in the 5--6 jet control sample. Additional checks at $100<\ptmiss\leq 200\GeV$
for both 5--6 jet events and 7-jet events yield consistent $\kappa$ values between the simulation and the data. These results,
as well as studies of the shape of the \nb distribution,
suggest that this discrepancy observed in the 2 \PQb jet bin at low \MJ is the result of a fluctuation.
Nevertheless, we assign systematic uncertainties to cover potential mismodeling of $\kappa$ as a function of \nb,
taking 10, 20, and 25\% as the uncertainties for events with $\nb=1$, $\nb=2$, and $\nb\geq 3$, respectively.

\subsection{Summary of systematic uncertainties in the background estimate}
\label{ssec:sys_summary}
Table~\ref{tab:systs:unc_summary} shows the total symmetrized systematic uncertainties in the $\kappa$ values used to compute the
background yields for each signal bin, based on the  uncertainties derived in the control samples. These uncertainties are obtained by combining the uncertainties
under the assumption of no correlation between any \njets, \nb, and \ptmiss dependence as follows,
\begin{linenomath}
\begin{equation}
\label{eq:systs:comb_syst}
\begin{aligned}
  \sigma^{\text{SR}}_{\text{low \ptmiss, j, b}}&=\sigma^{\text{5--6j}}_{\text{low \ptmiss, b}}\oplus\sigma^{2\ell}_{\text{low \ptmiss, j}},\\
  \sigma^{\text{SR}}_{\text{mid \ptmiss, j, b}}&=\sigma^{\text{5--6j}}_{\text{low \ptmiss, b}} \oplus \sigma^{2\ell}_{\text{low \ptmiss, j}}\oplus \sigma^{2\ell}_{\text{mid \ptmiss}},\\
  \sigma^{\text{SR}}_{\text{high \ptmiss, j, b}}&=\sigma^{\text{5--6j}}_{\text{low \ptmiss, b}} \oplus\sigma^{2\ell}_{\text{low \ptmiss, j}}\oplus \sigma^{2\ell}_{\text{high \ptmiss}},
\end{aligned}
\end{equation}
\end{linenomath}
where j and b are indices of the jet and \cPqb jet multiplicities, respectively. Here, $\sigma^{\text{5--6j}}_{\text{low \ptmiss, b}}$ refers to the uncertainty as a function of \nb derived in the low-\ptmiss bin of the single-lepton, 5--6 jet control sample; $\sigma^{2\ell}_{\text{low \ptmiss, j}}$ refers to the uncertainty as a function of \njets derived in the low-\ptmiss bin of the dilepton control sample;  and finally, $\sigma^{2\ell}_{\text{mid \ptmiss}}$ and $\sigma^{2\ell}_{\text{high \ptmiss}}$ refer to the uncertainty as a function of \ptmiss, integrated in \njets and \nb, derived in the dilepton control sample. Since the uncertainty as a function of \njets is derived in the low-\ptmiss bin of the dilepton sample, it already accounts for any mismodeling of the \ptmiss distribution at low \ptmiss, and thus no additional term is needed to account for such mismodeling in the first equation. Similarly, any mismodeling of the contribution of single-lepton mismeasured events at high \mT is already folded into the $\sigma^{\text{5--6j}}_{\text{low \ptmiss, b}}$ term, and thus no additional uncertainty is needed to account for this.

In practice, the three sources
of uncertainty listed above are implemented in the likelihood as six log-normal constraints.
A separate \lowmj and \highmj nuisance parameter is assigned for each of the quantities
$\sigma^{\text{SR}}_{\text{low \ptmiss, j, b}}$, $\sigma^{\text{SR}}_{\text{mid \ptmiss, j, b}}$, and
$\sigma^{\text{SR}}_{\text{high \ptmiss, j, b}}$. The \lowmj and \highmj nuisance parameters are decoupled,
based on the observation that the background contributions for which $\kappa>1$
have a \ptmiss dependence that is different at low \MJ and high \MJ. The total uncertainties with the
full Run 2 data set are in the range 13 to 39\%, increasing with \ptmiss.

\begin{table*}[btp!]
  \centering
  \topcaption{Systematic uncertainties in the background correction factors $\kappa$ associated with each signal bin based on the control sample studies described in Sections~\ref{ssec:cr2lveto} and \ref{ssec:cr56j} and combined according to Eq.(\ref{eq:systs:comb_syst}).\label{tab:systs:unc_summary}}
  \begin{scotch}{l ccc ccc ccc}
         & \multicolumn{3}{c}{$200<\ptmiss\leq350\GeV$} & \multicolumn{3}{c}{$350<\ptmiss\leq 500\GeV$} & \multicolumn{3}{c}{$\ptmiss>500\GeV$} \\
            & $\nb=1$ & $\nb=2$ & $\nb\geq3$& $\nb=1$ & $\nb=2$ & $\nb\geq3$& $\nb=1$ & $\nb=2$ & $\nb\geq3$\\
    \hline
    Low \MJ & 13\% & 22\% & 27\% & 20\% & 27\% & 31\% & 25\% & 30\% & 34\% \\
    High \MJ & 13\% & 22\% & 27\% & 22\% & 28\% & 32\% & 32\% & 36\% & 39\% \\
  \end{scotch}
\end{table*}

\section{Results and interpretation}
\label{sec:ResultsAndInterpretation}
\begin{figure*}[tbhp!]
\centering
\includegraphics[width=0.49\textwidth]{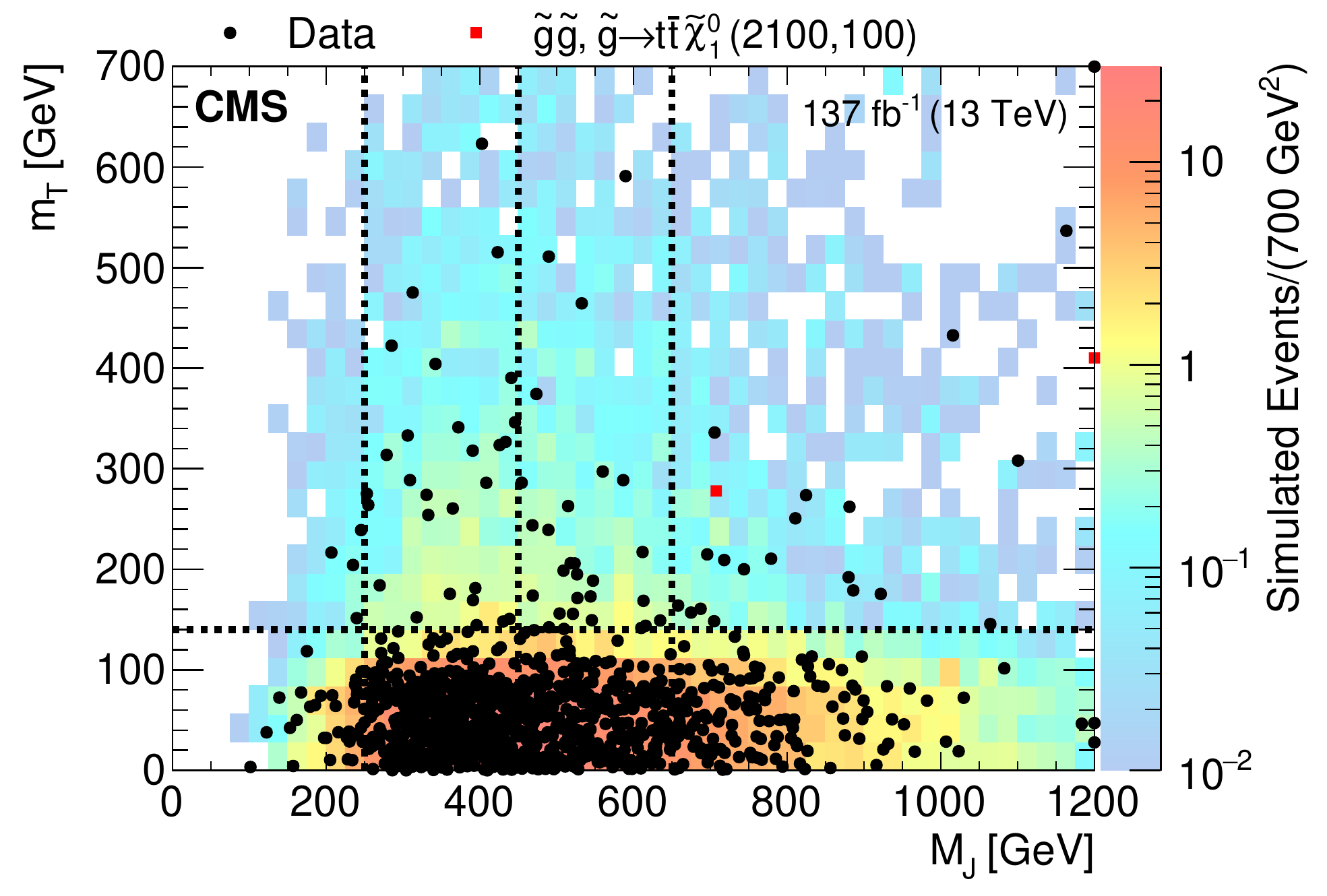}
\includegraphics[width=0.49\textwidth]{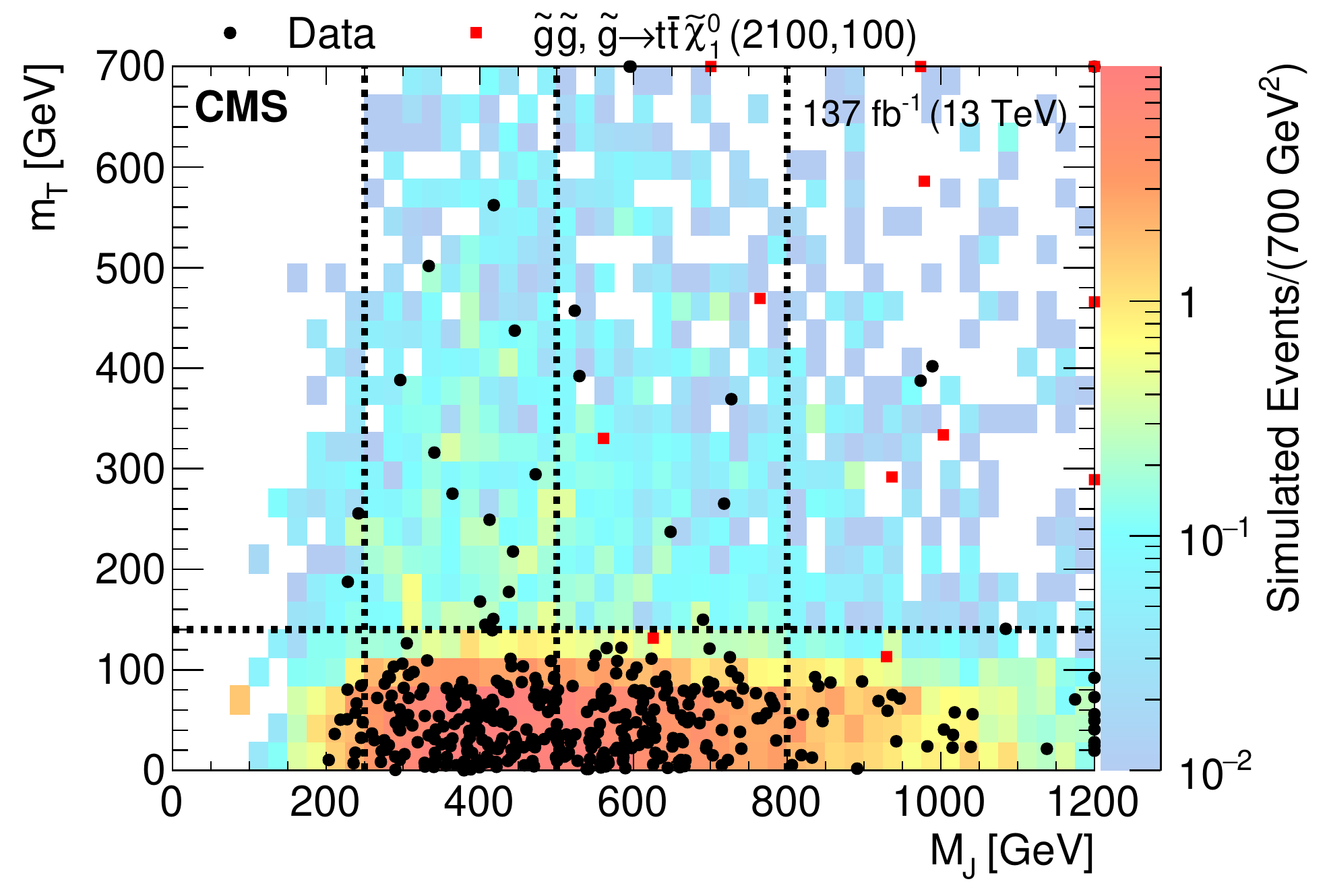}
\caption{Two-dimensional distributions in \MJ and \mT for both data and simulated event samples, integrated over the \njets and $\nb\geq2$ and shown separately
for the $350<\ptmiss\leq500\GeV$ bin (left) and the $\ptmiss> 500\GeV$ bin (right). The black dots represent events in data, the colored histogram shows the total
expected background yield per bin from simulation (not the actual predicted background),
and the red squares correspond to a representative random sample of signal events drawn from the simulated distribution for the
T1tttt model with $\mGlu=2100\GeV$ and $\mLSP=100\GeV$ for 137\fbinv.
Overflow events are shown on the edges of the plot.}
\label{fig:scatter}
\end{figure*}

Figure~\ref{fig:scatter} shows two-dimensional distributions of the data in the \MJ-\mT plane
after applying the baseline selection described in Section~\ref{sec:EvtSelAndRegions}, with separate plots for the intermediate- and high-\ptmiss bins. Both plots in the
figure are integrated over \njets and $\nb\geq2$ and hence do not represent the full sensitivity of the analysis.
Each event in data is represented by a single filled circle.
For comparison, the plots also show the expected total SM background based on simulation,
as well as an illustrative sample of the simulated signal distribution for the T1tttt model with $m(\PSg) = 2100\GeV$
and $m(\PSGczDo)=100\GeV$, plotted with one square per event, normalized to the same integrated luminosity as the data.
This model has a large mass splitting between the gluino and the neutralino, and
signal events typically have large values of \ptmiss. Qualitatively, the two-dimensional distribution of the data
corresponds well to the expected distribution for the SM background events. The highest \MJ, highest \ptmiss region shows
several simulated signal events for the T1tttt(2100, 100) model. However, only two observed events populate this region in the data.

\begin{figure*}[tbph!]
\centering
\includegraphics[width=0.48\textwidth]{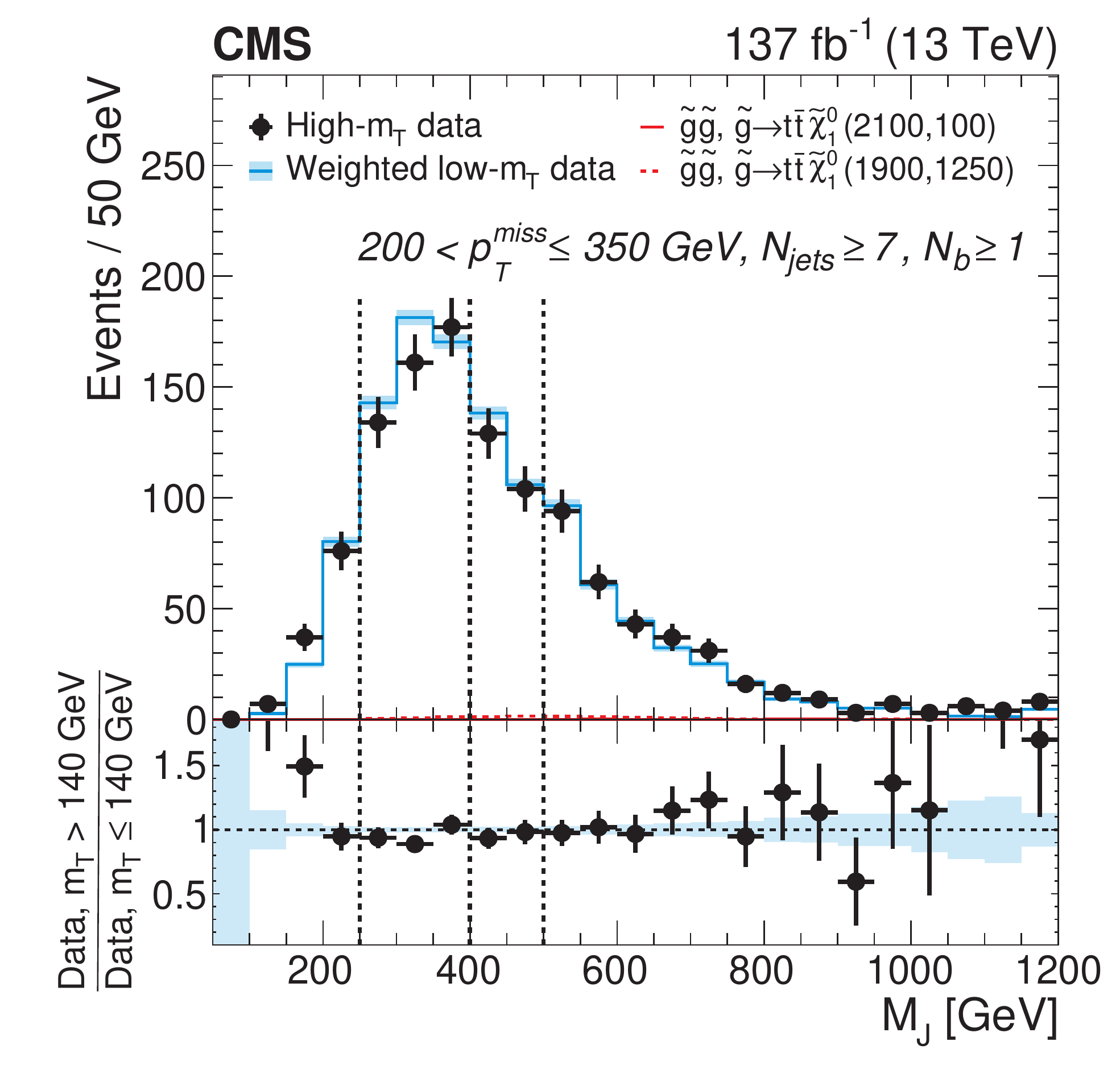}
\includegraphics[width=0.48\textwidth]{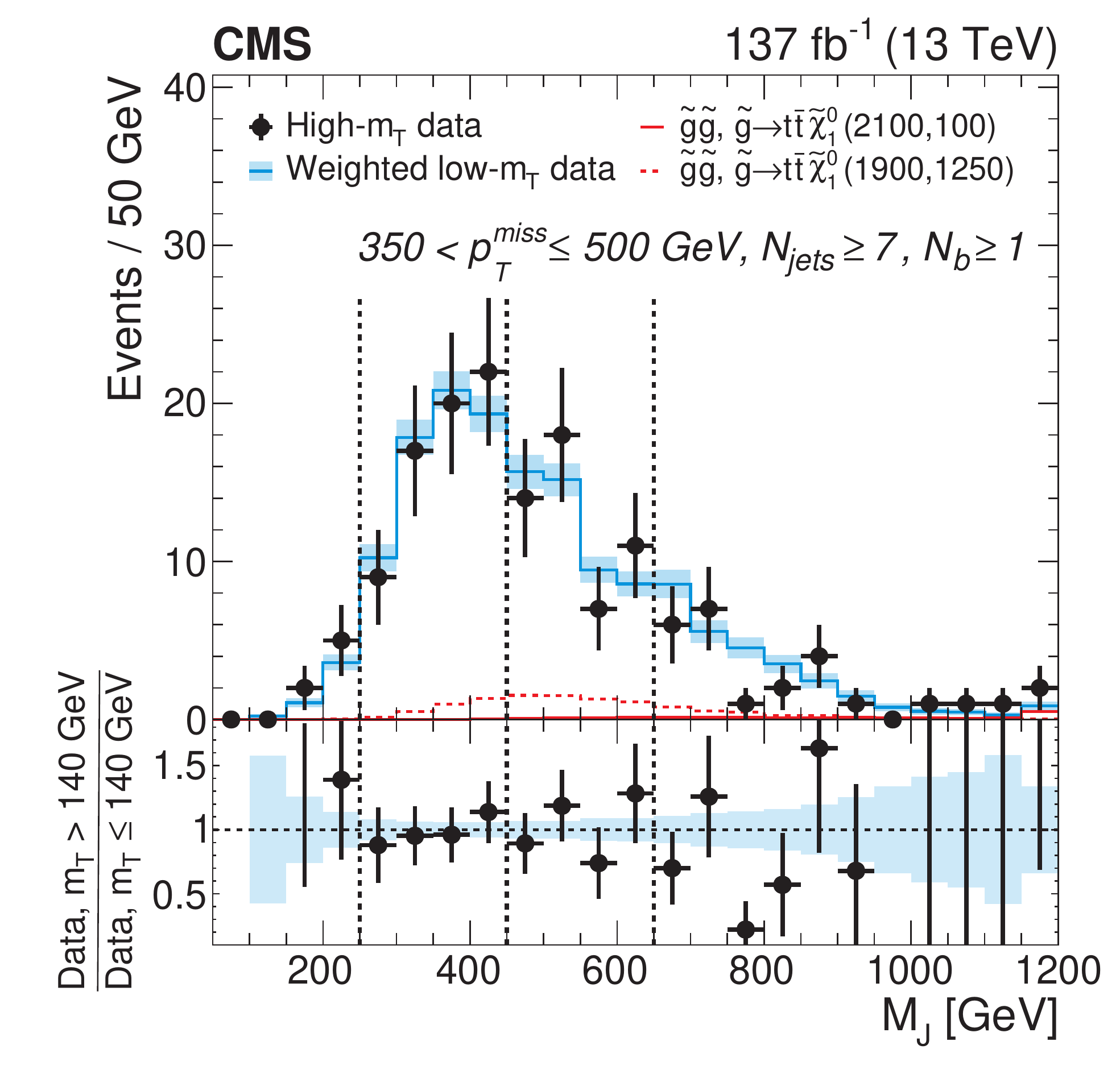}
\includegraphics[width=0.48\textwidth]{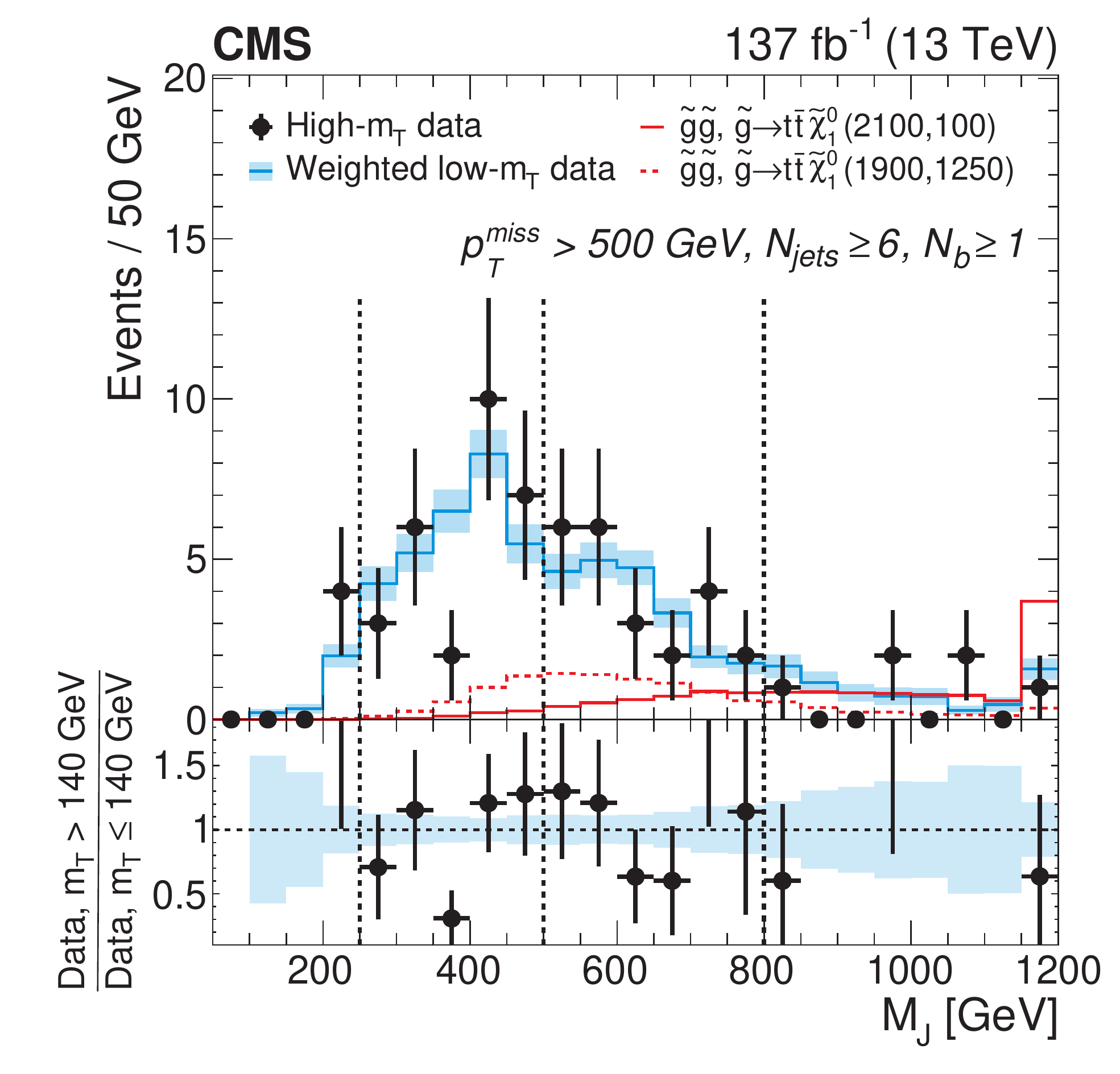}
\caption{Distributions of \MJ observed in data for $200<\ptmiss\leq 350\GeV$ (upper left), $350<\ptmiss\leq 500\GeV$
(upper right), and $\ptmiss > 500\GeV$ (lower) in the $1\ell$ data for low- and high-\mT regions.
In each plot, events in the R2A and R2B regions at low \mT have been weighted by the relevant $\kappa$ factor, and
the total low-\mT yield is normalized to the high-\mT yield to facilitate  comparison of the shapes of the distributions.
The vertical dashed line at $\MJ=250\GeV$ shows the lower boundary of regions R1 and R3, while the vertical
lines at higher \MJ values denote the lower \MJ boundaries of the signal regions R4A and R4B. The data
are integrated over the \nb and \njets signal bins. Two SUSY benchmark models are shown in the solid and
dashed red histograms.}
\label{fig:data2data_mj}
\end{figure*}

The basic principle of the analysis is illustrated in Fig.~\ref{fig:data2data_mj}, which
compares, in three separate \ptmiss regions, the \MJ distributions for low-\mT and high-\mT data.
The low-\mT data correspond to regions R1, R2A, and R2B. Here, each event in R2A or R2B is weighted with the
relevant $\kappa$ factor, and then the total low-\mT yield is normalized to the total
high-\mT yield in data. In the absence of signal, the shapes of
these distributions should be approximately consistent, as observed.
The low- and intermediate-\ptmiss regions (upper plots) show the background behavior with better
statistical precision, while the high-\ptmiss region (bottom) has a higher sensitivity to the signal.
For all three plots, integrals are performed over \njets and \nb, as indicated in the legends.

\begin{figure*}[tbph!]
\centering
\includegraphics[width=\textwidth]{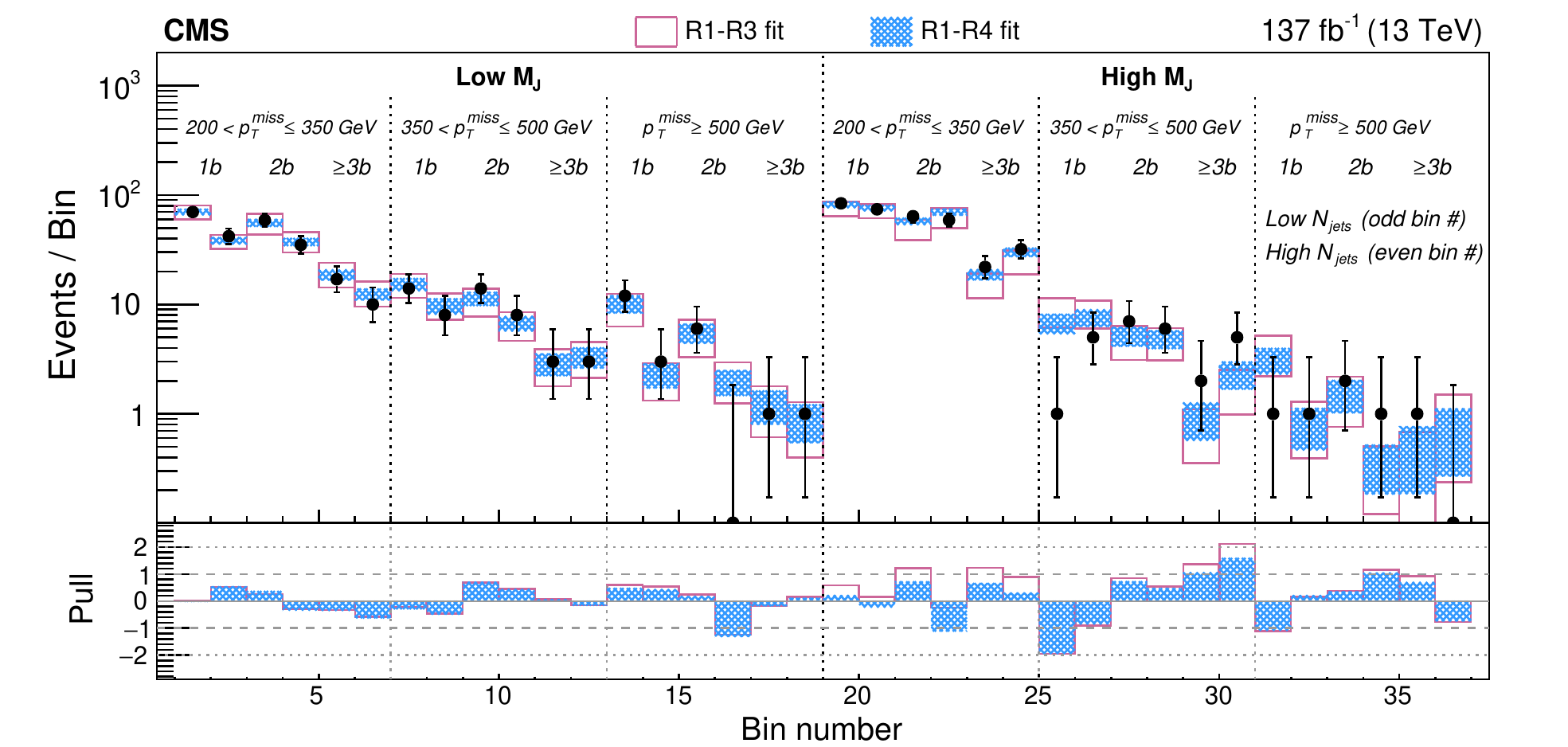}
\caption{Observed and predicted event yields in each signal region. The open rectangles represent the prediction and uncertainty obtained using event
yields from regions R1, R2, and R3 only (R1--R3 fit), while the hashed rectangles represent the prediction obtained when all regions are included in the fit (R1--R4 fit). The labels 1b, 2b, and $\geq$3b refer to $\nb=1$, $\nb=2$, and $\nb\geq3$ bins, respectively. In both cases, all statistical and systematic uncertainties are included. The bottom panel shows the pulls for both fits,
defined as $(N_\text{obs}-N_\text{pred})/\sqrt{\smash[b]{N_\text{pred}+(\sigma^\text{sys}_\text{pred})^2}}$. }
\label{fig:result}
\end{figure*}

Figure~\ref{fig:result} shows the observed event yields in all of the signal regions and bins of the analysis, the predicted backgrounds with their uncertainties obtained from the R1--R3 and R1--R4 fits,  and the pulls associated
with the fits. Both the R1--R3 and the R1--R4 fits are based on the null hypothesis, \ie, no
signal contribution.
We observe a broad pattern of consistency between the
observed data and predicted backgrounds in the search regions and bins.
Most of the pulls are less than one standard deviation (s.d.). The largest pull is $-2.0$ s.d.~and
occurs in the bin with $\MJ>650\GeV$, $350<\ptmiss\leq500\GeV$, $\njets=7$, and $\nb=1$.

Tables~\ref{tab:pred_results_lowmj} (low \MJ) and \ref{tab:pred_results_highmj} (high \MJ) present the same
information as in Fig.~\ref{fig:result}, but in detailed numerical form, including the observed and fitted yields in regions R1--R3, as well as the expected signal yields for the two T1tttt benchmark model points. Again,
the observed event yields in data are consistent with the background predictions.

\begin{table*}[tbph!]
  \centering
  \topcaption{Observed and predicted event yields for the signal regions (R4A) and
    background regions (R1--R3) in the \lowmj ABCD planes. For the R1--R3 fit, the values given for R1, R2A and R3 correspond to the observed yields in those regions.
    Expected yields for the two SUSY benchmark scenarios, T1tttt(2100, 100) and
    T1tttt(1900, 1250), are also given. The uncertainties in the prediction account for the amount of 
    data in the control samples, the precision of $\kappa$ from MC,
    and the systematic uncertainties in $\kappa$ assessed from control samples in data.}
  \label{tab:pred_results_lowmj}
  \cmsTable{
  \begin{scotch}{l cc r r r}
    $\mathcal{L}=137\fbinv$ & T1tttt(2100,100) & T1tttt(1900,1250) &R1--R3 fit & R1--R4 fit & Observed \\
    \hline
    \multicolumn{6}{c}{$200<\ptmiss\leq350\GeV$}  \\
    [\cmsTabSkip] 
    R1                               &       0.0&       1.1&     7706&   $7705 \pm 87$&      7706\\
    R2A: $\nb = 1$,  $\njets = 7$     &       0.0&       0.1&     1088&   $1088 \pm 32$&      1088\\
    R2A: $\nb = 1$, $\njets \geq 8$   &       0.0&       0.1&      732&    $736 \pm 26$&       732\\
    R2A: $\nb = 2$,  $\njets = 7$     &       0.0&       0.1&      879&    $882 \pm 30$&       879\\
    R2A: $\nb = 2$, $\njets \geq 8$   &       0.0&       0.3&      644&    $642 \pm 25$&       644\\
    R2A: $\nb \geq 3$,  $\njets = 7$  &       0.0&       0.2&      237&    $235 \pm 15$&       237\\
    R2A: $\nb \geq 3$, $\njets \geq 8$&       0.0&       0.5&      202&    $200 \pm 14$&       202\\
    R3                               &       0.0&       2.2&      472&    $473 \pm 20$&       472\\
    [\cmsTabSkip] 
    R4A: $\nb = 1$,  $\njets = 7$&       0.0&       0.2&     $70 \pm 10$&      $70.2 \pm 4.6$&        70\\
    R4A: $\nb = 1$, $\njets \geq 8$&       0.0&       0.3&      $37.7 \pm 5.6$&      $38.3 \pm 2.8$&        42\\
    R4A: $\nb = 2$,  $\njets = 7$&       0.0&       0.4&     $56 \pm 12$&      $55.7 \pm 4.5$&        59\\
    R4A: $\nb = 2$, $\njets \geq 8$&       0.0&       0.6&      $37.9 \pm 8.1$&      $37.4 \pm 3.1$&        35\\
    R4A: $\nb \geq 3$,  $\njets = 7$&       0.0&       0.4&      $19.2 \pm 4.9$&      $18.7 \pm 2.1$&        17\\
    R4A: $\nb \geq 3$, $\njets \geq 8$&       0.0&       0.9&      $12.9 \pm 3.3$&      $12.4 \pm 1.5$&        10\\
    [\cmsTabSkip] 
    \multicolumn{6}{c}{$350<\ptmiss\leq500\GeV$}  \\
    [\cmsTabSkip] 
    R1                               &       0.0&       0.9&       967&       $968 \pm 31$&        967\\
    R2A: $\nb = 1$,  $\njets = 7$     &       0.0&       0.1&       208&       $207 \pm 14$&        208\\
    R2A: $\nb = 1$, $\njets \geq 8$   &       0.0&       0.2&       150&       $148 \pm 12$&        150\\
    R2A: $\nb = 2$,  $\njets = 7$     &       0.0&       0.1&       139&       $142 \pm 12$&        139\\
    R2A: $\nb = 2$, $\njets \geq 8$   &       0.0&       0.3&       111&       $112 \pm 11$&        111\\
    R2A: $\nb \geq 3$,  $\njets = 7$  &       0.0&       0.2&        30&      $30.1 \pm 5.3$&        30\\
    R2A: $\nb \geq 3$, $\njets \geq 8$&       0.0&       0.6&        38&      $37.7 \pm 6.0$&        38\\
    R3                               &       0.1&       2.9&        68&      $67.0 \pm 6.5$&        68\\
    [\cmsTabSkip] 
    R4A: $\nb = 1$,  $\njets = 7$&       0.1&       0.3&      $15.2 \pm 3.7$&      $15.3 \pm 2.1$&        14\\
    R4A: $\nb = 1$, $\njets \geq 8$&       0.0&       0.4&       $9.9 \pm 2.7$&       $9.7 \pm 1.6$&         8\\
    R4A: $\nb = 2$,  $\njets = 7$&       0.1&       0.5&      $10.8 \pm 3.1$&      $11.3 \pm 1.7$&        14\\
    R4A: $\nb = 2$, $\njets \geq 8$&       0.1&       1.3&       $6.6 \pm 1.9$&       $6.8 \pm 1.1$&         8\\
    R4A: $\nb \geq 3$,  $\njets = 7$&       0.1&       0.7&       $2.8 \pm 1.1$&       $2.9 \pm 0.7$&         3\\
    R4A: $\nb \geq 3$, $\njets \geq 8$&       0.1&       2.1&       $3.3 \pm 1.2$&       $3.3 \pm 0.7$&         3\\
    [\cmsTabSkip] 
    \multicolumn{6}{c}{$\ptmiss> 500\GeV$}  \\
    [\cmsTabSkip] 
    R1                                      &       0.1&       0.6&      434&        $434 \pm 21$&       434\\
    R2A: $\nb = 1$,  $6\leq \njets \leq 7$   &       0.1&       0.1&      158&        $160 \pm 13$&       158\\
    R2A: $\nb = 1$, $\njets \geq 8$          &       0.0&       0.2&       41&      $41.7 \pm 6.4$&        41\\
    R2A: $\nb = 2$,  $6\leq \njets \leq 7$   &       0.1&       0.2&       80&      $80.5 \pm 8.8$&        80\\
    R2A: $\nb = 2$, $\njets \geq 8$          &       0.1&       0.3&       34&      $32.0 \pm 5.5$&        34\\
    R2A: $\nb \geq 3$,  $6\leq \njets \leq 7$&       0.1&       0.2&       20&      $19.8 \pm 4.5$&        20\\
    R2A: $\nb \geq 3$, $\njets \geq 8$       &       0.1&       0.5&       10&      $10.1 \pm 3.1$&        10\\
    R3                                      &       0.6&       3.2&       28&      $27.9 \pm 4.2$&        28\\
    [\cmsTabSkip] 
    R4A: $\nb = 1$,  $6\leq \njets \leq 7$&       0.6&       0.5&       $9.4 \pm 3.1$&      $10.2 \pm 1.9$&        12\\
    R4A: $\nb = 1$, $\njets \geq 8$&       0.3&       0.5&       $2.1 \pm 0.8$&       $2.3 \pm 0.6$&         3\\
    R4A: $\nb = 2$,  $6\leq \njets \leq 7$&       0.9&       1.0&       $5.3 \pm 2.0$&       $5.5 \pm 1.1$&         6\\
    R4A: $\nb = 2$, $\njets \geq 8$&       0.6&       1.3&       $2.1 \pm 0.9$&       $2.0 \pm 0.5$&         0\\
    R4A: $\nb \geq 3$,  $6\leq \njets \leq 7$&       0.8&       0.9&       $1.2 \pm 0.6$&       $1.2 \pm 0.4$&         1\\
    R4A: $\nb \geq 3$, $\njets \geq 8$&       0.8&       2.3&       $0.8 \pm 0.4$&       $0.9 \pm 0.3$&         1\\
 \end{scotch}
  }
\end{table*}

\begin{table*}[tbhp!]
  \centering
  \topcaption{Observed and predicted event yields for the signal regions (R4B) and
    background regions (R1--R3) in the \highmj ABCD planes. For the R1--R3 fit, the values given for R1, R2B and R3 correspond to the observed yields in those regions.
    Expected yields for the two SUSY benchmark scenarios, T1tttt(2100, 100) and
    T1tttt(1900, 1250), are also given. The uncertainties in the prediction account for the amount of 
    data in the control samples, the precision of $\kappa$ from MC,
    and the systematic uncertainties in $\kappa$ assessed from control samples in data.}
  \label{tab:pred_results_highmj}
  \cmsTable{
\begin{scotch}{l cc r r r}
    $\mathcal{L}=137\fbinv$ & T1tttt(2100,100) & T1tttt(1900,1250) &R1--R3 fit & R1--R4 fit & Observed \\
    \hline
    \multicolumn{6}{c}{$200<\ptmiss\leq350\GeV$}  \\
    [\cmsTabSkip] 
    R1                               &       0.0&       1.1&      7706&   $7705 \pm 87$&      7706\\
    R2B: $\nb = 1$,  $\njets = 7$     &       0.0&       0.1&       935&    $937 \pm 30$&       935\\
    R2B: $\nb = 1$, $\njets \geq 8$   &       0.1&       0.3&       961&    $959 \pm 30$&       961\\
    R2B: $\nb = 2$,  $\njets = 7$     &       0.0&       0.2&       600&    $606 \pm 24$&       600\\
    R2B: $\nb = 2$, $\njets \geq 8$   &       0.1&       0.6&       832&    $821 \pm 28$&       832\\
    R2B: $\nb \geq 3$,  $\njets = 7$  &       0.0&       0.2&       168&    $171 \pm 13$&       168\\
    R2B: $\nb \geq 3$, $\njets \geq 8$&       0.1&       1.1&       306&    $308 \pm 17$&       306\\
    R3                               &       0.0&       2.2&       472&    $473 \pm 20$&       472\\
    [\cmsTabSkip] 
    R4B: $\nb = 1$,  $\njets = 7$&       0.1&       0.2&     $76 \pm 11$&      $81.7 \pm 5.6$&        84\\
    R4B: $\nb = 1$, $\njets \geq 8$&       0.2&       0.5&     $72 \pm 10$&      $76.3 \pm 4.9$&        74\\
    R4B: $\nb = 2$,  $\njets = 7$&       0.2&       0.4&     $49 \pm 10$&      $57.6 \pm 4.3$&        64\\
    R4B: $\nb = 2$, $\njets \geq 8$&       0.3&       1.5&     $63 \pm 13$&      $70.0 \pm 5.1$&        59\\
    R4B: $\nb \geq 3$,  $\njets = 7$&       0.1&       0.6&      $15.2 \pm 3.9$&      $18.8 \pm 2.1$&        22\\
    R4B: $\nb \geq 3$, $\njets \geq 8$&       0.4&       2.6&      $24.9 \pm 6.1$&      $30.1 \pm 2.9$&        32\\
    [\cmsTabSkip] 
    \multicolumn{6}{c}{$350<\ptmiss\leq500\GeV$}  \\
    [\cmsTabSkip] 
    R1                               &       0.0&       0.9&       967&        $968 \pm 31$&       967\\
    R2B: $\nb = 1$,  $\njets = 7$     &       0.0&       0.0&        78&      $72.2 \pm 8.2$&        78\\
    R2B: $\nb = 1$, $\njets \geq 8$   &       0.1&       0.1&        95&      $92.4 \pm 9.5$&        95\\
    R2B: $\nb = 2$,  $\njets = 7$     &       0.1&       0.0&        54&      $55.8 \pm 7.3$&        54\\
    R2B: $\nb = 2$, $\njets \geq 8$   &       0.1&       0.2&        65&      $66.1 \pm 8.1$&        65\\
    R2B: $\nb \geq 3$,  $\njets = 7$  &       0.0&       0.1&         8&       $9.1 \pm 2.9$&         8\\
    R2B: $\nb \geq 3$, $\njets \geq 8$&       0.1&       0.4&        16&      $18.7 \pm 4.2$&        16\\
    R3                               &       0.1&       2.9&        68&      $67.0 \pm 6.5$&        68\\
    [\cmsTabSkip] 
    R4B: $\nb = 1$,  $\njets = 7$&       0.2&       0.1&       $8.7 \pm 2.6$&       $6.8 \pm 1.4$&         1\\
    R4B: $\nb = 1$, $\njets \geq 8$&       0.2&       0.3&       $8.4 \pm 2.4$&       $7.6 \pm 1.4$&         5\\
    R4B: $\nb = 2$,  $\njets = 7$&       0.2&       0.1&       $4.7 \pm 1.6$&       $5.2 \pm 1.0$&         7\\
    R4B: $\nb = 2$, $\njets \geq 8$&       0.4&       0.7&       $4.6 \pm 1.5$&       $4.9 \pm 0.9$&         6\\
    R4B: $\nb \geq 3$,  $\njets = 7$&       0.2&       0.1&       $0.7 \pm 0.4$&       $0.9 \pm 0.3$&         2\\
    R4B: $\nb \geq 3$, $\njets \geq 8$&       0.5&       1.3&       $1.8 \pm 0.8$&       $2.3 \pm 0.7$&         5\\
    [\cmsTabSkip] 
    \multicolumn{6}{c}{$\ptmiss> 500\GeV$}  \\
    [\cmsTabSkip] 
    R1                                      &       0.1&       0.6&        434&       $434 \pm 21$&        434\\
    R2B: $\nb = 1$,  $6\leq \njets \leq 7$   &       0.1&       0.0&         49&      $46.9 \pm 7.0$&        49\\
    R2B: $\nb = 1$, $\njets \geq 8$          &       0.2&       0.1&         13&      $13.2 \pm 3.7$&        13\\
    R2B: $\nb = 2$,  $6\leq \njets \leq 7$   &       0.2&       0.0&         18&      $18.5 \pm 4.3$&        18\\
    R2B: $\nb = 2$, $\njets \geq 8$          &       0.3&       0.2&          7&       $7.6 \pm 2.8$&         7\\
    R2B: $\nb \geq 3$,  $6\leq \njets \leq 7$&       0.2&       0.0&          4&       $4.5 \pm 2.1$&         4\\
    R2B: $\nb \geq 3$, $\njets \geq 8$       &       0.4&       0.3&          5&       $4.3 \pm 2.0$&         5\\
    R3                                      &       0.6&       3.2&         28&      $27.9 \pm 4.2$&        28\\
    [\cmsTabSkip] 
    R4B: $\nb = 1$,  $6\leq \njets \leq 7$&       1.0&       0.1&       $3.7 \pm 1.5$&       $3.1 \pm 0.9$&         1\\
    R4B: $\nb = 1$, $\njets \geq 8$&       1.1&       0.3&       $0.8 \pm 0.4$&       $0.8 \pm 0.3$&         1\\
    R4B: $\nb = 2$,  $6\leq \njets \leq 7$&       1.4&       0.1&       $1.5 \pm 0.7$&       $1.5 \pm 0.5$&         2\\
    R4B: $\nb = 2$, $\njets \geq 8$&       2.0&       0.6&       $0.3 \pm 0.2$&       $0.4 \pm 0.2$&         1\\
    R4B: $\nb \geq 3$,  $6\leq \njets \leq 7$&       1.1&       0.1&       $0.4 \pm 0.3$&       $0.5 \pm 0.3$&         1\\
    R4B: $\nb \geq 3$, $\njets \geq 8$&       2.4&       1.0&       $0.9 \pm 0.6$&       $0.7 \pm 0.4$&         0\\
 \end{scotch}
  }
\end{table*}

The results are first interpreted in terms of the simplified model T1tttt
of SUSY particle production.
This model is characterized by just two mass parameters, \mGlu and \mLSP.
To determine which sets of masses, or mass points, are excluded by the data,
we generate a set of simulated signal samples in which the mass parameters are varied across the range to which
the analysis is potentially sensitive. These samples are used to determine the number of events that would be expected at
each mass point, given the theoretical production cross section for this point. To assess which model points
can be excluded by the data, it is necessary to evaluate
the systematic uncertainties associated with the expected number of observed signal events.

Systematic uncertainties in the expected signal yields
account for uncertainties in the trigger, lepton identification, jet identification, and \PQb tagging
efficiencies in simulated events; uncertainties in the distributions of \ptmiss,
number of pileup vertices, and ISR jet multiplicity; and uncertainties in the jet energy
corrections, renormalization and factorization scales, and integrated luminosity~\cite{CMS-PAS-LUM-17-001,CMS-PAS-LUM-17-004,CMS-PAS-LUM-18-002}. Each systematic uncertainty is evaluated in each of the analysis bins separately, and the
uncertainties are treated as symmetric log-normal distributions. In the case that the sizes of up and down variations
are not the same, the variation having larger absolute value is taken. If the sign of variations
changes bin-by-bin, the correlation between bins is preserved, while the value that has the larger
absolute variation is taken. A summary of the magnitude of the uncertainty due to each systematic source
across sensitive signal bins for each of the two signal benchmark points is shown in Table~\ref{tab:unc:sig}.

\begin{table*}[tbph!]\centering
\topcaption{Range of values for the systematic  uncertainties in the signal efficiency and acceptance across sensitive bins, specifically across high \ptmiss signal bins for T1tttt(2100,100) and high \njets signal bins for T1tttt(1900,1250). Uncertainties due to a particular source are
treated as fully correlated among bins, while uncertainties due to different sources are treated as uncorrelated.}
\label{tab:unc:sig}
\renewcommand{\arraystretch}{1.2}
\begin{scotch}{lcc}
\multirow{2}{*}{Source}&\multicolumn{2}{c}{Relative uncertainty [\%]}\\
& \multicolumn{1}{c}{T1tttt(2100,100)} & \multicolumn{1}{c}{T1tttt(1900,1250)} \\
  \hline
MC sample size                 & 3--8   & 7--15   \\
Renormalization and factorization scales           & 1--2   & 2--4   \\
Fast MC~\ptmiss resolution     & 1--2   & 1--5 \\
Lepton efficiency              & 7--9   & 4--5 \\
Trigger efficiency             & 1      & 1   \\
\PQb tagging efficiency        & 2--8   & 2--8  \\
Mistag efficiency              & 1      & 1--3  \\
Jet energy corrections         & 1--5   & 2--11\\
Initial-state radiation        & 1--7   & 1--10 \\
Jet identification             & 1      & 1    \\
Pileup                         & 1--2   & 1--4 \\
Integrated luminosity          & 2.3--2.5 & 2.3--2.5  \\
\end{scotch}
\end{table*}

An upper limit on the production cross section at 95\% confidence level (\CL) is estimated using the modified
frequentist \CLs method~\cite{Junk:1999kv,0954-3899-28-10-313,CMS-NOTE-2011-005},
with a one-sided profile likelihood ratio test statistic, using an asymptotic approximation~\cite{Cowan:2010js}.
The statistical uncertainties from data counts in the
control regions are modeled by Poisson terms. All systematic uncertainties
are multiplicative and are treated as log-normal distributions.  Exclusion limits are also estimated
for $\pm1$ s.d.~variations on the production cross section based on the approximate NNLO+NNLL
calculation.

\begin{figure}[tbph!]
  \centering
  \includegraphics[width=0.49\textwidth]{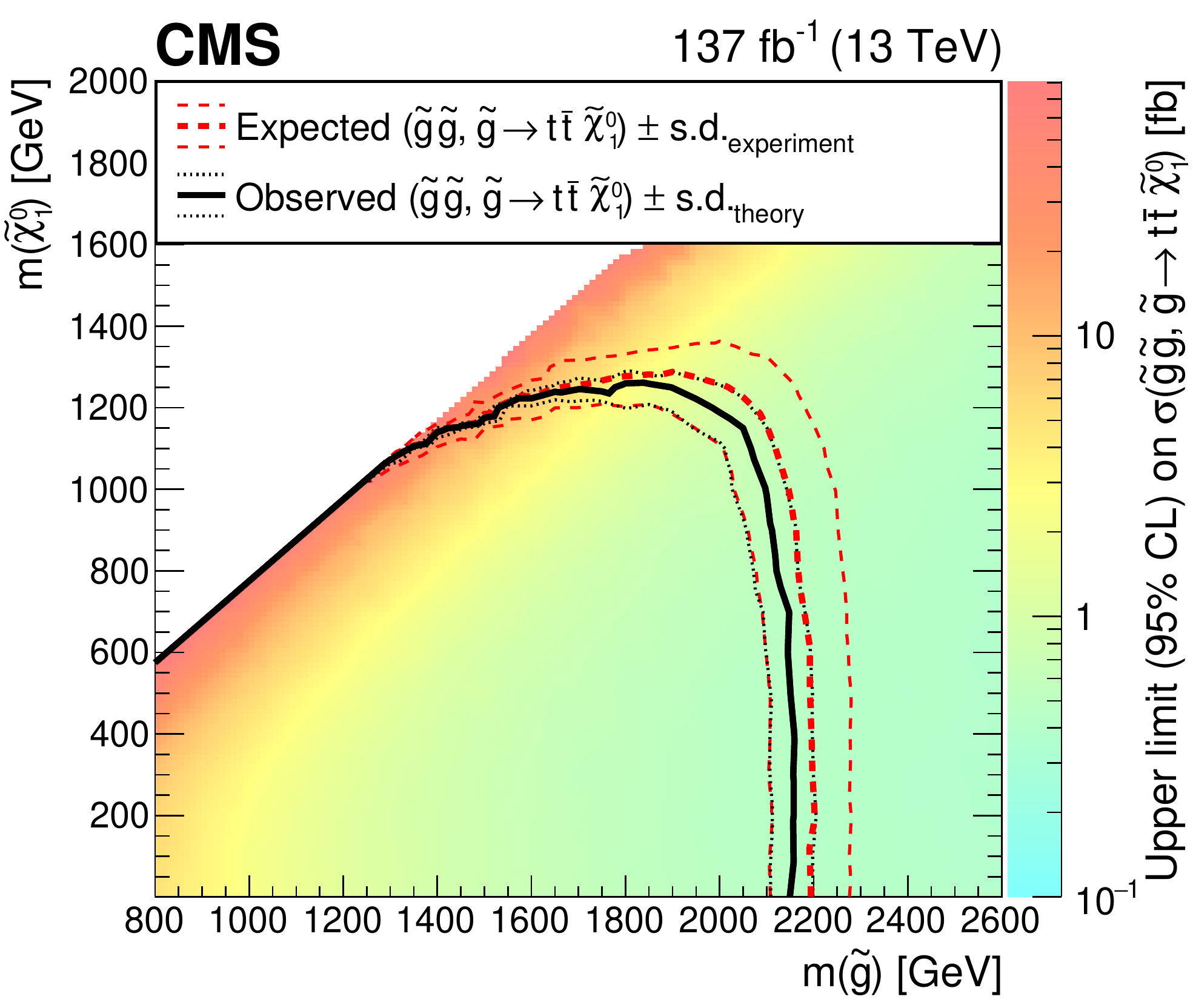}
  \caption{Interpretation of the results in the T1tttt model. The colored regions
    show the upper limits (95\% \CL) on the production cross section for
    $\Pp\Pp\to\sGlu\sGlu,\sGlu\to\ttbar\PSGczDo$ in the
    \mGlu-\mLSP plane. The curves show the expected and observed limits on
    the corresponding SUSY particle masses obtained by comparing the excluded
    cross section with theoretical cross sections.}
  \label{fig:t1tttt_limits_scan}
\end{figure}

\begin{figure}[tbph!]
  \centering
  \includegraphics[width=0.49\textwidth]{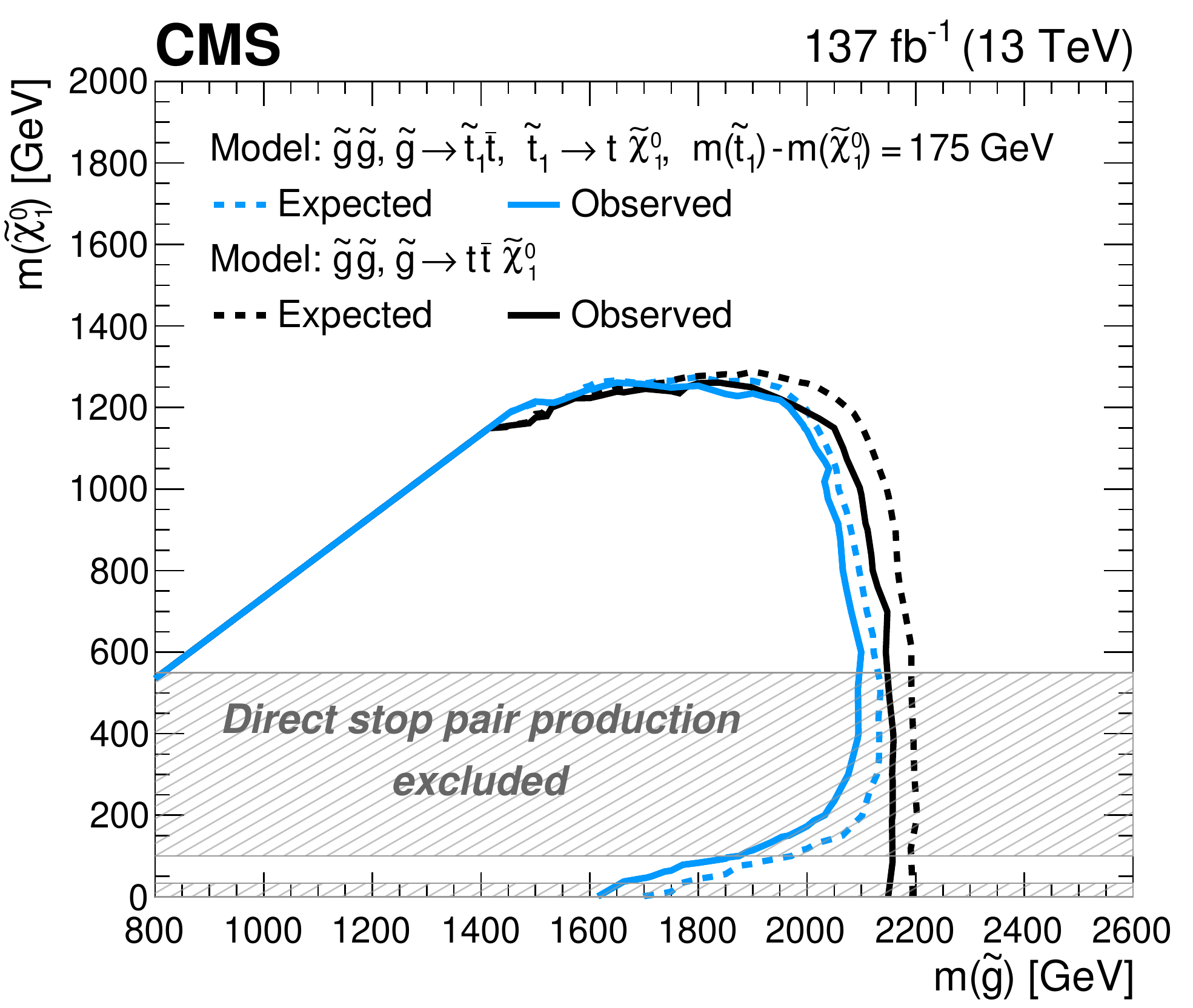}
  \caption{Interpretation of the results in the T5tttt model. The expected and observed upper limits do not take into account contributions from direct top squark pair production; however, its effect for $m(\PSGczDo) > 550\GeV$ is small. The T1tttt interpretation results are also shown for comparison. }
  \label{fig:t5tttt_limits_scan}
\end{figure}

Figures~\ref{fig:t1tttt_limits_scan} and \ref{fig:t5tttt_limits_scan} show the corresponding excluded cross section  regions at 95\% \CL for the T1tttt and T5tttt models, respectively,
in the \mGlu-\mLSP plane. These regions correspond
to excluded cross sections under the assumption that the branching fraction for
the given process is 100\%.
For T1tttt, gluinos with masses of up to approximately 2150\GeV are excluded for \PSGczDo masses
up to about 700\GeV.
The highest limit on the \PSGczDo mass is approximately 1250\GeV, attained for
\mGlu of about 1700--1900\GeV. The observed limits for T1tttt are within the
$1\sigma$ uncertainty of the expected limits across the full scan range.

The T5tttt model allows us to extend the interpretation of the results to scenarios in which the top squark
is lighter than the gluino. Rather than considering a large set of models with
independently varying top squark masses, we consider the extreme case in which the
top squark has approximately the smallest mass consistent with two-body decay,
$m(\PSQt) \approx m(\cPqt) + m(\PSGczDo)$, for a range of gluino and neutralino masses. The decay
kinematics for such extreme, compressed mass spectrum models correspond to the lowest
signal efficiency for given values of $m(\sGlu)$ and $m(\PSGczDo)$, because the top
quark and the \PSGczDo are produced at rest in the top squark frame. As a consequence,
the excluded signal cross section for fixed values of $m(\sGlu)$ and $m(\PSGczDo)$ and
with $m(\sGlu) > m(\PSQt_1) \ge m(\PQt) +  m(\PSGczDo)$ is minimized for this extreme
model point. In particular, at low \mLSP the neutralino carries
very little momentum, thus reducing the value of \mT, and resulting in significantly
lower sensitivity for T5tttt than T1tttt. In this kinematic region, the
sensitivity to the signal is in fact dominated by the events that have at least two
leptonic \PW boson decays, which produce additional \ptmiss, as well as a tail
in the \mT distribution. Although such dilepton events are nominally excluded
in the analysis, a significant number of these signal events escape the dilepton
veto.

For physical consistency, the signal model used in the T5tttt study should
include not only gluino pair production, but also direct top squark pair production,
$\PSQt\PASQt$, referred to as T2tt. For $m(\PSGczDo) <33\GeV$ and $100 < m(\PSGczDo) < 550\GeV$, with $m(\PSQt) - m(\PSGczDo) = 175\GeV$, the T2tt model is excluded in direct searches for $\PSQt\PASQt$ production~\cite{CMS-SUS-18-003,Sirunyan:2019ctn}. For $33 < m(\PSGczDo) < 100\GeV$, the T2tt model is not excluded due to the difficulty in assessing the rapidly changing acceptance with the finite event count available in simulation. We have verified that for $m(\PSGczDo) > 550\GeV$, where the T2tt model remains unexcluded, adding the
contribution from the T2tt process to our analysis regions does not have a significant effect on the sensitivity.
For simplicity, in Fig.~\ref{fig:t5tttt_limits_scan}, we have
based the exclusion curve on T5tttt only, without including the additional
T2tt process.

As with all SUSY particle mass limits obtained in the context of simplified models, it
is important to recognize that the results can be significantly weakened
if the assumptions of the model fail to hold. In particular, the presence
of alternative decay modes could reduce the number of expected events
for the given selection. However, cross
section limits remain valid if they are interpreted as limits on
cross section multiplied by the branching fraction for the assumed decay mode.

\section{Summary}
\label{sec:Summary}
A search is performed for an excess event yield above that expected for standard model
processes using a data sample of proton-proton collision events with an integrated luminosity of 137\fbinv at
$\sqrt{s} = 13\TeV$. The experimental signature is characterized by a single isolated
lepton, multiple jets, at least one \PQb-tagged jet, and large missing transverse momentum.
No significant excesses above the standard model backgrounds
are observed.  The results are interpreted in the framework of simplified models that describe natural
supersymmetry scenarios.  For gluino pair production followed by the three-body decay
$\PSg\to\ttbar\PSGczDo$ (T1tttt model), gluinos with masses below about 2150\GeV are excluded at 95\% confidence
level for neutralino masses up to 700\GeV. The highest excluded neutralino mass is about 1250\GeV.
For the two-body gluino decay $\PSg\to\PSQt_1\cPaqt$ with
$\PSQt_1\to\cPqt\PSGczDo$ (T5tttt model), the results are generally similar, except at low neutralino masses,
where the excluded gluino mass is somewhat lower.
These results extend previous gluino mass limits~\cite{Sirunyan:2017fsj} from this search by about
250\GeV, due to both the data sample increase and the analysis reoptimization enabled by it.
These mass limits are among the most stringent constraints on this supersymmetry model to date.

\begin{acknowledgments}
We congratulate our colleagues in the CERN accelerator departments for the excellent performance of the LHC and thank the technical and administrative staffs at CERN and at other CMS institutes for their contributions to the success of the CMS effort. In addition, we gratefully acknowledge the computing centres and personnel of the Worldwide LHC Computing Grid for delivering so effectively the computing infrastructure essential to our analyses. Finally, we acknowledge the enduring support for the construction and operation of the LHC and the CMS detector provided by the following funding agencies: BMBWF and FWF (Austria); FNRS and FWO (Belgium); CNPq, CAPES, FAPERJ, FAPERGS, and FAPESP (Brazil); MES (Bulgaria); CERN; CAS, MoST, and NSFC (China); COLCIENCIAS (Colombia); MSES and CSF (Croatia); RPF (Cyprus); SENESCYT (Ecuador); MoER, ERC IUT, PUT and ERDF (Estonia); Academy of Finland, MEC, and HIP (Finland); CEA and CNRS/IN2P3 (France); BMBF, DFG, and HGF (Germany); GSRT (Greece); NKFIA (Hungary); DAE and DST (India); IPM (Iran); SFI (Ireland); INFN (Italy); MSIP and NRF (Republic of Korea); MES (Latvia); LAS (Lithuania); MOE and UM (Malaysia); BUAP, CINVESTAV, CONACYT, LNS, SEP, and UASLP-FAI (Mexico); MOS (Montenegro); MBIE (New Zealand); PAEC (Pakistan); MSHE and NSC (Poland); FCT (Portugal); JINR (Dubna); MON, RosAtom, RAS, RFBR, and NRC KI (Russia); MESTD (Serbia); SEIDI, CPAN, PCTI, and FEDER (Spain); MOSTR (Sri Lanka); Swiss Funding Agencies (Switzerland); MST (Taipei); ThEPCenter, IPST, STAR, and NSTDA (Thailand); TUBITAK and TAEK (Turkey); NASU (Ukraine); STFC (United Kingdom); DOE and NSF (USA).

\hyphenation{Rachada-pisek} Individuals have received support from the Marie-Curie programme and the European Research Council and Horizon 2020 Grant, contract Nos.\ 675440, 752730, and 765710 (European Union); the Leventis Foundation; the A.P.\ Sloan Foundation; the Alexander von Humboldt Foundation; the Belgian Federal Science Policy Office; the Fonds pour la Formation \`a la Recherche dans l'Industrie et dans l'Agriculture (FRIA-Belgium); the Agentschap voor Innovatie door Wetenschap en Technologie (IWT-Belgium); the F.R.S.-FNRS and FWO (Belgium) under the ``Excellence of Science -- EOS" -- be.h project n.\ 30820817; the Beijing Municipal Science \& Technology Commission, No. Z181100004218003; the Ministry of Education, Youth and Sports (MEYS) of the Czech Republic; the Deutsche Forschungsgemeinschaft (DFG) under Germany’s Excellence Strategy – EXC 2121 ``Quantum Universe" -- 390833306; the Lend\"ulet (``Momentum") Programme and the J\'anos Bolyai Research Scholarship of the Hungarian Academy of Sciences, the New National Excellence Program \'UNKP, the NKFIA research grants 123842, 123959, 124845, 124850, 125105, 128713, 128786, and 129058 (Hungary); the Council of Science and Industrial Research, India; the HOMING PLUS programme of the Foundation for Polish Science, cofinanced from European Union, Regional Development Fund, the Mobility Plus programme of the Ministry of Science and Higher Education, the National Science Center (Poland), contracts Harmonia 2014/14/M/ST2/00428, Opus 2014/13/B/ST2/02543, 2014/15/B/ST2/03998, and 2015/19/B/ST2/02861, Sonata-bis 2012/07/E/ST2/01406; the National Priorities Research Program by Qatar National Research Fund; the Ministry of Science and Education, grant no. 3.2989.2017 (Russia); the Programa Estatal de Fomento de la Investigaci{\'o}n Cient{\'i}fica y T{\'e}cnica de Excelencia Mar\'{\i}a de Maeztu, grant MDM-2015-0509 and the Programa Severo Ochoa del Principado de Asturias; the Thalis and Aristeia programmes cofinanced by EU-ESF and the Greek NSRF; the Rachadapisek Sompot Fund for Postdoctoral Fellowship, Chulalongkorn University and the Chulalongkorn Academic into Its 2nd Century Project Advancement Project (Thailand); the Nvidia Corporation; the Welch Foundation, contract C-1845; and the Weston Havens Foundation (USA).
\end{acknowledgments}
\bibliography{auto_generated}
\cleardoublepage \appendix\section{The CMS Collaboration \label{app:collab}}\begin{sloppypar}\hyphenpenalty=5000\widowpenalty=500\clubpenalty=5000\vskip\cmsinstskip
\textbf{Yerevan Physics Institute, Yerevan, Armenia}\\*[0pt]
A.M.~Sirunyan$^{\textrm{\dag}}$, A.~Tumasyan
\vskip\cmsinstskip
\textbf{Institut f\"{u}r Hochenergiephysik, Wien, Austria}\\*[0pt]
W.~Adam, F.~Ambrogi, T.~Bergauer, M.~Dragicevic, J.~Er\"{o}, A.~Escalante~Del~Valle, M.~Flechl, R.~Fr\"{u}hwirth\cmsAuthorMark{1}, M.~Jeitler\cmsAuthorMark{1}, N.~Krammer, I.~Kr\"{a}tschmer, D.~Liko, T.~Madlener, I.~Mikulec, N.~Rad, J.~Schieck\cmsAuthorMark{1}, R.~Sch\"{o}fbeck, M.~Spanring, W.~Waltenberger, C.-E.~Wulz\cmsAuthorMark{1}, M.~Zarucki
% ...
\vskip\cmsinstskip
\dag: Deceased\\
1:  Also at Vienna University of Technology, Vienna, Austria\\
% ...
\end{sloppypar}
\end{document}